\documentclass[sigconf]{acmart}

\usepackage[utf8]{inputenc}
\usepackage[inline]{enumitem}
\usepackage{multirow}
\usepackage{bbding}
\usepackage{makecell}
\usepackage{comment}
\usepackage{subcaption}
\usepackage{graphicx}  
\usepackage{array}

\usepackage{colortbl}
\usepackage{xcolor}
\AtBeginDocument{%
  }

\copyrightyear{2026}
\acmYear{2026}
\setcopyright{cc}
\setcctype{by}
\acmConference[CHI '26]{Proceedings of the 2026 CHI Conference on Human Factors in Computing Systems}{April 13--17, 2026}{Barcelona, Spain}
\acmBooktitle{Proceedings of the 2026 CHI Conference on Human Factors in Computing Systems (CHI '26), April 13--17, 2026, Barcelona, Spain}
\acmPrice{}
\acmDOI{10.1145/3772318.3791209}
\acmISBN{979-8-4007-2278-3/2026/04}

\definecolor{darkgreen}{rgb}{0,0.5,0}
\definecolor{orange}{rgb}{1,0.5,0}
\definecolor{teal}{rgb}{0,0.5,0.5}
\definecolor{darkpurple}{rgb}{0.5, 0, 0.5}
\definecolor{olive}{rgb}{0.6,0.6,0}

\newcommand {\rr}[1]{{{#1}\normalfont}}
\newcommand {\cam}[1]{{{#1}\normalfont}}

\newcommand {\toc}[1]{{{#1}\normalfont}}

\newcommand{\interface}[0]{{{\it Interface}}}
\newcommand{\highlight}[0]{{{\it HighlightInterface}}}
\newcommand{\level}[0]{{{\it HighlightLevel}}}
\newcommand{\plain}[0]{{{\sc PLAIN}}}
\newcommand{\aoc}[0]{{{\sc AOC}}}
\newcommand{\rake}[0]{{{\sc RAKE}}}
\newcommand{\gptsm}[0]{{{\sc GP-TSM}}}
\newcommand{\as}[0]{{{\sc SUMMARY}}}

\newcommand{\combined}[0]{{{\sc Combined}}}

\newcommand{\easy}[0]{{{\sc Easy}}}
\newcommand{\hard}[0]{{{\sc Hard}}}

\newcommand{\zero}[0]{{{\sc Level 0}}}
\newcommand{\one}[0]{{{\sc Level 1}}}

\newcommand{\inlinequote}[1]{{{\it ``#1''}}}

\newcommand{\ontext}[0]{{\small {\it On Text}}}
\newcommand{\offtext}[0]{{\small {\it Off Text}}}
\newcommand{\sustained}[0]{{\small {\it Sustained Reading}}}
\newcommand{\hopping}[0]{{\small {\it Hopping}}}
\newcommand{\local}[0]{{\small {\it Local}}}
\newcommand{\distal}[0]{{\small {\it Distal}}}
\newcommand{\fc}[0]{{\small {\it Fixation Count}}}
\newcommand{\ttt}[0]{{\small {\it Total Fixation}}}
\newcommand{\gd}[0]{{\small {\it Gaze Duration}}}
\newcommand{\ffd}[0]{{\small {\it First Fixation Duration}}}
\newcommand{\regressionin}[0]{{\small {\it RegressionInline}}}
\newcommand{\regressionbt}[0]{{\small {\it RegressionBetweenLine}}}

\newcommand{\anovazh}[5]{{($F_{#1, #2} = #3$, $p #4$, \eff{#5})}} 
\newcommand{\eff}[1]{$\eta_p^2 = #1$}

\newcommand{\kruskalsse}[4]{{($H_{#1} = #2$, $p < .001$, \effwk{#4})}} 
\newcommand{\kruskalnse}[4]{{($H_{#1} = #2$, $p = #3$, \effwk{#4})}}
\newcommand{\effwk}[1]{$\eta^2 = #1$} 

\newcommand{\mannwhitneyu}[3]{{($U=#1$, $Z=#2$, $p<0.001$, $r=#3$)}}

\newcommand{\meanse}[2]{($M = #1$, $SE = #2$)}

\begin{document}






\title[Desirable Unfamiliarity]{Desirable Unfamiliarity: Insights from Eye Movements on Engagement and Readability of Dictation Interfaces}
\author{Zhaohui Liang}
\authornote{Both authors contributed equally to this research.}

\affiliation{%
  \department{School of Creative Media,}
  \institution{City University of Hong Kong}
  \city{Hong Kong}
  \country{China}
}
\email{zhaohuileon@gmail.com}
\author{Yonglin Chen}
\authornotemark[1]

\affiliation{%
\department{School of Creative Media,}
  \institution{City University of Hong Kong}
  \institution{Southern University of Science and Technology}
  \city{Hong Kong}
  \country{China}
}
\email{yonglin0711@gmail.com}

\author{Naser Al Madi}
\affiliation{%
\department{Department of Computer Science,}
  \institution{Colby College}
  \city{Waterville}
  \state{Maine}
  \country{USA}}
\email{nsalmadi@colby.edu}

\author{Can Liu}
\affiliation{%
\department{School of Creative Media,}
  \institution{City University of Hong Kong}
  \city{Hong Kong}
  \country{China}
}
\email{canliu@cityu.edu.hk}
\authornote{Corresponding Author.}





\renewcommand{\shortauthors}{Liang and Chen, et al.}

\begin{abstract}
  Transcripts displayed on dictation interfaces can be hard to read due to recognition errors and disfluencies. LLM-based text auto-correction could help, but changing the text during production could lead to distraction and unintended phrasing. To understand how to balance readability, attention, and accuracy, we conducted an eye-tracking experiment with 20 participants to compare five dictation interfaces: PLAIN (real-time transcription), AOC (periodic corrections), RAKE (keyword highlights), GP-TSM (grammar-preserving highlights), and SUMMARY (LLM-generated abstractive summary). By analyzing participants’ gaze patterns during speech composition and reviewing processes, we found that during composition, participants spent only 7\%-11\% of their time in active reading regardless of the interface. Although SUMMARY introduced unfamiliar words and phrasing during composition, it was easier to read and more preferred by participants. Our findings suggest a high user tolerance for altering spoken words in LLM-enabled diction interfaces. 

\end{abstract}





\begin{teaserfigure}
\centering
  \includegraphics[width=.95\textwidth]{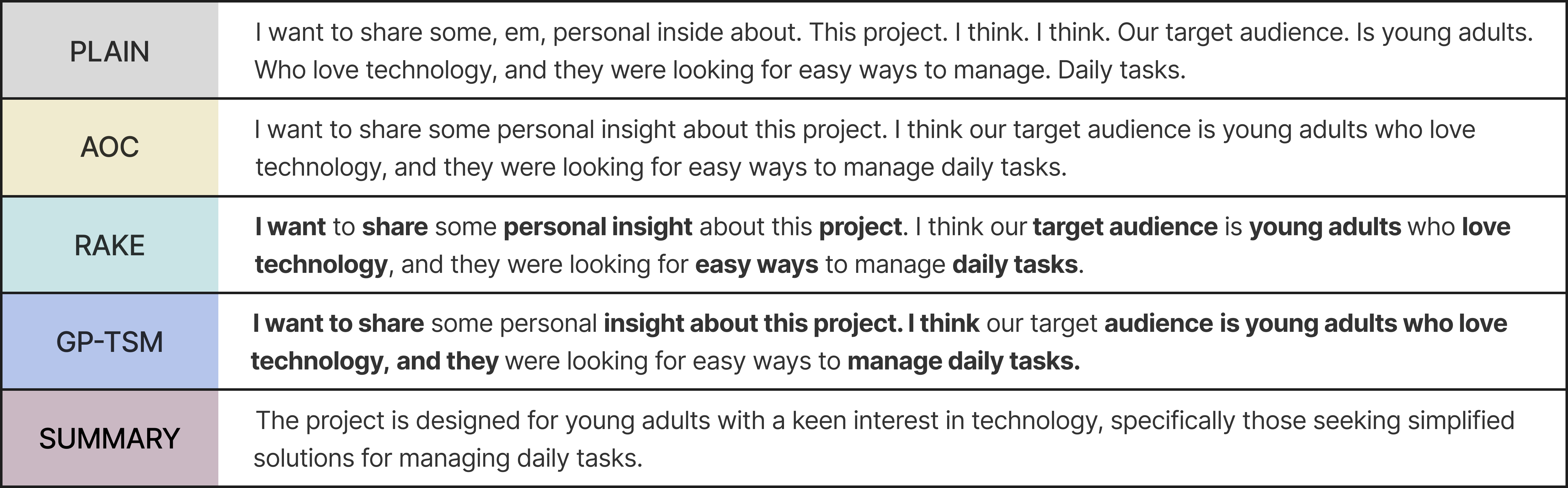}
  \caption{Illustrating the five dictation interfaces compared in our study: PLAIN (real-time transcription), AOC (periodic corrections), RAKE (keywords highlighting), GP-TSM (grammar-preserving highlighting), and SUMMARY (LLM-generated abstractive summary).}
  \label{fig:teaser}
\end{teaserfigure}


\maketitle


\section{Introduction}

Speech-to-text (STT) technology has become ubiquitous, following the breakthroughs of Automatic Speech Recognition (ASR) and Natural Language Processing (NLP). 
Most operating systems today support STT input, which turns every text editor into a dictation interface. Mobile phone users increasingly use STT input for writing memos. Dictation software is increasingly used in personal, professional and educational settings, where being able to quickly review the spoken content is essential. 
Despite its widespread availability, it is known that raw transcripts can be hard to read and process due to disfluencies, recognition errors, excessive punctuation, and potentially disorganization in long composition \cite{10.1145/3571884.3597134}. These errors cannot be completely avoided in real-time transcripts, even with the latest speech recognition technology~\cite{liao2023improving, lou2020end, tanaka2018neural}. Contextual correction has been employed in products such as Google Voice Typing\footnote{research.google/blog/an-all-neural-on-device-speech-recognizer} and Apple Notes\footnote{www.icloud.com/notes}, which immediately display the words from raw transcription as interim results and correct them at the user's pause. Recent dictation products such as Wispr Flow\footnote{https://wisprflow.ai/} leverage AI to remove fillers and correct spelling before displaying the transcribed text. However, there are trade-offs: changing the displayed text during production could cause visual distraction to the user, while displaying text after production (post-processing) could compromise responsiveness. Therefore, dictation interfaces need to balance readability and responsiveness when displaying spoken text.

Beyond the word- or phrase- level errors caused by disfluency and pronunciation, the inherent impromptu nature of speech input produces verbose and disorganized content in longform composition, which is another reason why transcripts are hard to read \cite{zayats19_interspeech, lin2024rambler}. 
Previous research shows faster "memory decay of speech production and a preference towards gist abstraction in memory traces" compared to typing or handwriting~\cite{10.1145/3571884.3597134}. 
Building on this, a recent dictation tool called Rambler\cite{lin2024rambler}, designed for composing longform text, advocates for ``gist-based'' interfaces that aid the review of spoken text with \rr{two methods: highlighting keywords and LLM-generated summaries. }
For extracting keywords, there are classic methods based on word frequencies \cite{aizawa2003information} or recent deep learning methods \cite{issa2023comparative}. Concerns about LLM summaries altering the original meaning of the text led to recent work introducing Grammar-Preserving Text Saliency Modulation (GP-TSM) \cite{gu2024ai}, which is shown to be helpful for skimming through longform text by highlighting the core sentence structures.  

\rr{However, the effectiveness of these above solutions on readability has not been evaluated.}
Current text-altering solutions to help users read spoken text range from minor spelling correction to rewritten summaries. LLM-based text correction or summaries may alter the text to an extent the users would not agree with. So where should we draw the line? Would users prefer reading the raw transcript for maximum responsiveness and fidelity, or post-processed text for better readability? How much tolerance do users have for their original spoken text being altered, while speaking and reviewing the text? 
\rr{For aiding the review of longform spoken text, which method is more effective}, extractive methods (word highlighting) or abstractive methods (summaries)? In addition, we also want to find out which word-highlighting technique works better in this context.

In this work, we aim to answer the questions above by conducting an eye-tracking study to investigate users' reading behavior and engagement across representative dictation interfaces that differ in how spoken text is processed and displayed. Although eye movement research has studied reading extensively~\cite{rayner1998eye, torrance2016reading, commarford2004models, shadiev2014investigating}, the use of eye movement in the evaluation of dictation interfaces remains unexplored in the literature. In our study, we analyzed participants' eye movement data during \emph{speech production} and \emph{reviewing} processes respectively, when using each interface. These interfaces were chosen as they differ in: \emph{fidelity to original speech}, \emph{methods for aiding review}, \emph{interface responsiveness} 
and \emph{stability}. The five dictation interfaces we chose are \rr{listed in Table \ref{tab:five_intertfaces}: Plain (raw transcript), AOC (Accumulative Offline Correction - raw transcript in realtime with periodical updates of corrected text), RAKE (AOC with standard keyword highlighting), GP-TSM (AOC with keyword highlighting that preserves sentence structure~\cite{gu2024ai}), and SUMMARY (LLM-reworded to be consise). } More details of each interface can be found in Section~\ref{interfaces}. 


\definecolor{gray1}{gray}{0.9}
\definecolor{gray2}{gray}{0.8}
\definecolor{gray3}{gray}{0.6}

\begin{table*}
\centering

\setlength{\extrarowheight}{5pt} 

\caption{Dimensions where the five dictation interfaces differ.}
\label{tab:five_intertfaces}

\begin{tabular}{|m{2.3cm}|m{2.0cm}|m{2.5cm}|m{2.8cm}|m{2.8cm}|m{2.5cm}|}
\hline
& \textbf{PLAIN} & \textbf{AOC} & \textbf{RAKE} & \textbf{GP-TSM} & \textbf{SUMMARY} \\
\hline

Fidelity to \newline original speech & 
As spoken &
\cellcolor{gray1}Minor correction &
\cellcolor{gray1}Minor correction &
\cellcolor{gray1}Minor correction &
\cellcolor{gray3}Rewritten \\
\hline

Aid to review & 
No &
\cellcolor{gray1}Smooth wording &
\cellcolor{gray3}Smooth wording \newline + Extractive highlight &
\cellcolor{gray3}Smooth wording \newline + Extractive highlight &
\cellcolor{gray3}Abstractive gist\\
\hline

Interface \newline responsiveness & 
Immediate &
\cellcolor{gray1}Immediate \newline + 10s Update &
\cellcolor{gray1}Immediate \newline + 10s Update &
\cellcolor{gray1}Immediate \newline + 10s Update &
\cellcolor{gray1}Immediate \newline + 10s Update \\
\hline

Interface stability & 
No change &
\cellcolor{gray1}Word change &
\cellcolor{gray2}Word change \newline + highlight &
\cellcolor{gray2}Word change \newline + highlight &
\cellcolor{gray3}Sentence change \\
\hline
\end{tabular}

\end{table*}

    
    
    
    


We recruited 20 university students to compose social media posts of around 250 to 300 words using the five dictation interfaces while their eye-movement was recorded during speech production and a subsequent phase where participants reviewed the text. We analyzed their visual engagement with the text, the reading effort and gaze patterns affected by each interface, as well as their subjective user experience. 

\toc{Our findings reveal that participants spent only 7-11\% of their time reading the text during speech production regardless of the interface, highlighting the cognitive demands of composing spoken content. We also found that \as{}, despite introducing unfamiliar words, significantly reduced reading difficulty and was more preferred by participants. \rr{We term this phenomenon ``Desirable Unfamiliarity'': In the context of displaying self-spoken text, an observed tendency that LLM-generated abstractive summaries lead to enhanced readability and comprehension despite introducing unfamiliar words.}
The simple classic highlighting strategy (\rake{}) was surprisingly more effective in guiding eye movement compared to \gptsm{}. These insights challenge existing assumptions about dictation interfaces and provide actionable implications for future design.} 

\section{Related work}


To provide more context for our contributions, we discuss three areas of highly related work: Dictation Interfaces, Eye Movement in Reading and Production, and Information extraction from text.

\subsection{Dictation Interfaces}

Modern dictation tools enable efficient text input. A wide range of commercial tools and models support speech-to-text transcription, which can be categorized into real-time and offline transcription. 
Real-time transcription allows users to view transcribed text instantly, with latency as low as milliseconds, making it widely used in scenarios such as voice input, simultaneous interpretation, and subtitle generation \cite{kumar2012voice, xiong2019dutongchuan, che2017automatic}. Offline transcription, on the other hand, processes complete audio files to generate transcriptions and is commonly applied in social media platforms and meeting-minutes generation \cite{modha2020detecting, rennard2023abstractive}.

Although speaking is faster than typing or writing, it contains errors caused by speech artifacts, such as disfluencies and repetitions \cite{zebrowski1991duration, howell1995automatic}. Additionally, transcribed text is often lengthy, making it difficult to review important information, which increases the burden of reading \cite{bokhove2018automated}. Moreover, both transcription methods have inherent limitations. Real-time transcription provides fast response times but at the expense of reduced accuracy \cite{hain2011transcribing}, while offline transcription achieves higher accuracy but lacks the ability to deliver real-time feedback to users \cite{gaskell2016speech}.
Modern speech and NLP research has made much progress in cleaning disfluency and speech recognition errors, mostly from post-processing \cite{liao2023improving, lou2020end, tanaka2018neural}. Moreover, with the support of large language models (LLMs), the processing of transcribed text has become more flexible and adaptive. Advanced GPT-4o multi-model LLM can process and respond to real-time audio input in as little as 232 milliseconds, which is similar to human response time in a conversation, and can observe tone, multiple speakers, or even background noise\footnote{https://openai.com/index/hello-gpt-4o/}.

Our paper introduces a method that combines real-time transcription with offline transcription post-processing, utilizing approaches similar to GPT-4o. This method applies offline transcription techniques and various NLP processing methods to real-time transcription, enabling low-latency corrections of the transcribed text. By doing so, we examine how different text correction approaches influence the user experience and preferences during the process of producing and reviewing text using dictation interfaces.


\subsection{Eye Movement in Reading and Production}
Eye movement during reading has been extensively studied, where saccades (jumps between words) and fixations (the duration for which the eyes remain stationary on a word) constitute the primary eye movements during reading \cite{rayner1998eye}.  Despite the numerous eye-tracking studies on reading, we found no studies that focused on analyzing gaze patterns in the evaluation of dictation interfaces. Understanding these gaze patterns can help improve the design of dictation interfaces and better understand the needs of users during production and reviewing. Here, we review relevant research on reading during writing composition as a closely related task to text composition through speech.

As we consider how to improve dictation interfaces as a whole, we need to be mindful of the two main tasks users perform: Speech production (speaking) and reviewing produced text.  The two tasks differ in purpose, and hence it is likely that different cognitive processes are employed in monitoring during speaking and in thorough reading for comprehension during reviewing.  Previous research shows that eye movement patterns differ according to reading purpose and characteristics \cite{torrance2016reading, strukelj2018one}. In fact, during written text production (composition) it was found that writers spent only 5.8\% of their time in sustained reading, compared to 72.3\% of their time during reading unfamiliar text \cite{torrance2016reading}. 

In written text production, previous research shows that writers read and evaluate their writing at the word level as words are produced, this \emph{local} reading behavior is associated with error-detection processes and facilitates word-level planning \cite{beers2010adolescent}.  At the same time, writers often reread previous sections of their developing text to ``evaluate and improve their sentence construction'' \cite{beers2010adolescent}.  This \emph{distal} reading pattern far from the word being produced ``may allow writers to compose texts with greater overall coherence in accordance with their rhetorical goals'' \cite{beers2010adolescent}.

At the word level, word familiarity is another aspect that is important in the context of dictation interfaces considering that during reviewing users read familiar text that they produced. Reading familiar text is likely to make word identification easier \cite{leroy2014effect}, and that would be reflected in shorter fixation duration and higher probability of skipping short predictable words \cite{rayner1998eye, torrance2016reading}. Moreover, word identification is a stepping stone to comprehension, where difficulties often result in re-reading text through jumps back called regressions \cite{rayner2006eye, sharmin2016reading}. The presence of these regressions with a particular dictation interface can indicate difficulties in reading produced text at the comprehension level.


Our paper provides a detailed description of eye movement while using different dictation interfaces during speaking and reviewing produced text. In addition to understanding how users visually engage with spoken text, this permits us to understand how elements of word familiarity and text readability play a role in forming users' perceptions and preferences of dictation interfaces.

\subsection{Information Extraction from text}

Extensive research has explored text information extraction methods for enhancing readability, supporting skimming or reducing reading cost, which can be divided into extractive methods and abstractive methods. Extractive methods combine a set of text segments from the original document(s) to form a summary, and a variety of modeling \cite{dong2018banditsum, jadhav2018extractive, jia2020neural, luo2019reading, xu2015extractive} and learning \cite{zhang2018neural, narayan2018ranking} approaches have been explored in this context. Abstractive summarization, which generates novel sentences to capture the essence of the content \cite{allahyari2017text, zhang2017sentence, zhong2022unsupervised}, offers greater flexibility, conciseness, and readability.

Extractive summarization is often used in conjunction with text saliency modulation or highlighting, a well-researched technique that enhances readability by adjusting visual attributes of the text to guide attention and emphasize key information. QuickSkim \cite{10.1145/1753846.1754093} reduces opacity for non-essential words when skimming is detected, while other systems adjust font size or weight to emphasize key information. Strobelt et al. tested nine common highlighting techniques, but most studies focus on tasks like visual search rather than reading performance and skimming \cite{strobelt2015guidelines}. 
While extractive summarization traditionally outperformed in terms of accuracy and coherence, recent advancements in natural language generation with LLMs have significantly enhanced the effectiveness and popularity of abstractive summarization \cite{widyassari2022review}. Laskar et al. employ LLMs to add summaries to transcribed text, facilitating quick recall of spoken content, which is commonly used in online meeting scenarios \cite{laskar2023building}. Rambler utilizes LLMs to support users' diverse summarizing needs for transcribed text, including semantic segmenting and merging \cite{lin2024rambler}.

Our research studies two distinct extractive summarization methods: keyword highlighting (\rake{}) and grammar-preserving key component highlighting (\gptsm{}), as well as an abstractive method using LLMs (\as{}). We apply these techniques to users' transcribed text (self-produced and verbose) to evaluate their effectiveness and explore user preferences in the context of transcription.



\section{Experiment Platform}

\begin{figure*}
    \centering
    \includegraphics[width=1\linewidth]{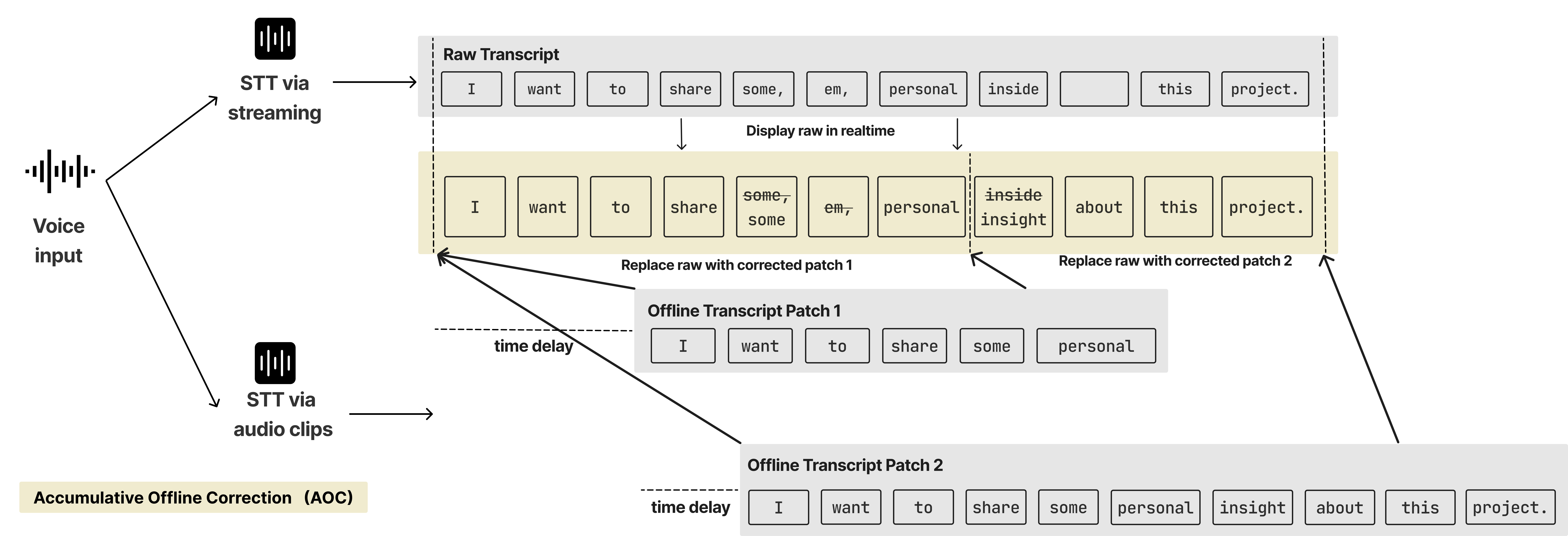}
    \caption{Illustration of the mechanism of Accumulative Offline Correction(AOC). The interface displays raw transcripts in real-time, while the text gets periodically replaced by corrected text. }
    \label{fig:AOC}
\end{figure*}

To investigate how to enhance the user experience with Speech-to-text (STT) interfaces and reduce the reading cost of spoken text, we built an experimental platform that allows the experimenter to choose between five dictation interfaces. We introduce them one by one as follows. 

\subsection{Five Dictation Interfaces} \label{interfaces}
\paragraph{\textbf{\plain{} --- Real-time Transcription Baseline}}
We use AssemblyAI’s Streaming Speech-to-Text (STT) service\footnote{www.assemblyai.com} as the baseline for real-time transcription. We utilized a WebSocket API to stream audio data, which provided initial transcripts within a few hundred milliseconds. As additional context became available, the system progressively refined the transcripts, enhancing their accuracy over time.

\paragraph{\textbf{\aoc{} --- Accumulative Offline Correction}}
Plain real-time transcription is fast but often prone to transcription errors, punctuation mistakes, and reduced readability. Non-real-time transcription methods, which take input as entire audio files, can yield better transcription accuracy by applying contextual correction. Yet such methods require post-processing of the audio, and cannot provide a real-time transcribing experience. How about combining the advantages of both methods? 
We introduce the Accumulative Offline Correction technique (AOC) - a new transcription method to provide better recognition accuracy while preserving real-time feedback. Our approach is to display raw transcripts in real-time, then periodically replace parts of the earlier transcripts with offline-transcribed text. As shown in \ref{fig:AOC}, we implemented two virtual recorders in the system: Recorder 1 for plain real-time transcription and Recorder 2 for offline transcription. At each time interval, a recording segment from Recorder 2 is extracted and sent to AssemblyAI’s Speech Recognition API, which allows users to upload an audio file offline and receive the transcription after a delay.

\paragraph{\textbf{\rake{} --- Keywords highlighting}}
We use Rapid Automatic Keyword Extraction (RAKE) method
\cite{rose2010automatic}\footnote{https://pypi.org/project/rake-nltk/} to detect keywords in the user's transcript. \rake{} is an unsupervised, domain-independent, and language-independent method for extracting keywords from individual texts. Compared to using LLM for keyword detection, \rake{} detects keywords with minimal latency. The raw transcript goes through the same correction process as \aoc{} and is then sent to the RAKE component to detect keywords. Similar to the \aoc{} condition, this interface displays raw transcript users speak, then periodically replaces text with corrected \aoc{} text and keywords highlighted in bold. 

\paragraph{\textbf{\gptsm{} --- Extractive Summary}}

We implemented extractive summary inspired by an AI-Resilient Text Rendering Technique called Grammar-Preserving Text Saliency Modulation (GP-TSM) \cite{gu2024ai}. GP-TSM compresses sentences by LLM to de-emphasize less essential words while preserving grammatical structure. While the original version of GP-TSM provided three-level highlights, we only provided two levels: highlighted and non-highlighted. This adjustment was made to facilitate the comparison with \rake{} and focus on understanding what content to highlight. Unlike the static nature of GP-TSM's fixed text, speech transcription text is dynamic. We employ a zero-shot chain of thought prompting approach to achieve accumulated identification of key segments, ensuring that each highlight iteration builds on previous results, with minimal changes to previously highlighted sections. On a text of approximately 200 words, the original GP-TSM requires around 40 seconds of processing time, while our simplified version takes less than 5 seconds, enabling it to perform within shorter time intervals. \gptsm{} processes and renders text based on AOC results, also first displays the raw transcript and replaces it.

\paragraph{\textbf{\as{} --- Abstractive Summary}}
For the abstractive summary, we use OpenAI's LLM API to generate reconstructed text, which is approximately 80\% of the original text length and automatically corrects any grammatical errors in the user's transcript. \as{} processes and renders text based on \aoc{} results. This interface directly displays the generated summary with a delay, instead of first displaying raw transcripts, unlike \aoc{}, \rake{} and \gptsm{} as explained above. 

\subsection{System Implementation}
We implemented the Dictation system with Python Flask and Vue. We access the OpenAI API to use the gpt-4-turbo-2024-04-09 model for all LLM use. The system runs on a Linux server, and users use the system on a Windows web browser. 

\paragraph{\textbf{\rr{API Latency and Interface Update Interval}.}} Since AssemblyAI’s Speech Recognition API takes around 6 seconds to process an offline audio file, and our system also includes \rake{} (2 seconds delay), \gptsm{} (4 seconds delay), and \as{} (4 seconds delay) processing, we set the interval between sending patches to be 10 seconds for \aoc{}, \rake{}, \gptsm{}, and \as{}, which was chosen after our tests on all four interfaces. This choice is partially limited by the API delays, and partially made to balance the responsiveness and stability of the interface. Overly frequent changes on the interface are distracting for users.

\paragraph{\textbf{Robustness.}} We use Zero-shot Chain-of-Thought (CoT) and Diversity Sampling to enhance the robustness of \gptsm{} and \as{}. The Zero-shot CoT prompt is provided in the attachment. \rr{Our pilot experiments revealed that unstable LLM output causes significant fluctuations in output density, ranging from excessive highlighting to overly-brief summaries, which can make the visual representation quite distracting. To mitigate this instability, we employ Diversity Sampling and CoT to improve the consistency of output quality.} When sending LLM requests, the system sends eight concurrent identical requests and selects the response with the closest length to their average (mean cumulative highlight length for \gptsm{}, mean summary length for \as{}). \cam{This helped remove extreme cases and stabilize the text representation.}  

\paragraph{\textbf{User Interface.}} Users can adjust different modes through the settings area at the top of the interface, then click "Record" to start recording and stop button to end the recording. For the two LLM-based modes, \gptsm{} and \as{}, we have added a refresh button, allowing users to re-render the text after the recording ends to avoid any unstable results from the LLM affecting the experiment.
After the recording ends, the system generates timestamps and text end positions during the user's speaking process to support subsequent analysis in combination with eye-tracking data.
\section{Experiment Design}
We designed a within-subject experiment with one independent variable \interface{} of five values: [\plain, \aoc, \rake, \gptsm, \as]. By comparing \plain{} and \aoc{}, we aim to evaluate the effects of periodical correction of raw transcripts. Then by comparing \rake, \gptsm, and \as, we want to understand the differences between these representative approaches to extracting information from transcripts. 
We aim to answer the following research questions:
\begin{itemize}
    \item [RQ1] How do participants visually engage with the text during speech production and reviewing processes?
    \item [RQ2] How do interfaces affect visual engagement with text during production?
    \item [RQ3] How do interfaces affect reading effort and behavior during reviewing?
    \item [RQ4] Which interface(s) do participants prefer?
\end{itemize}

\begin{figure*}
  \centering
  \begin{subfigure}[b]{0.25\textwidth}
  	\centering
  	\includegraphics[width=\textwidth]{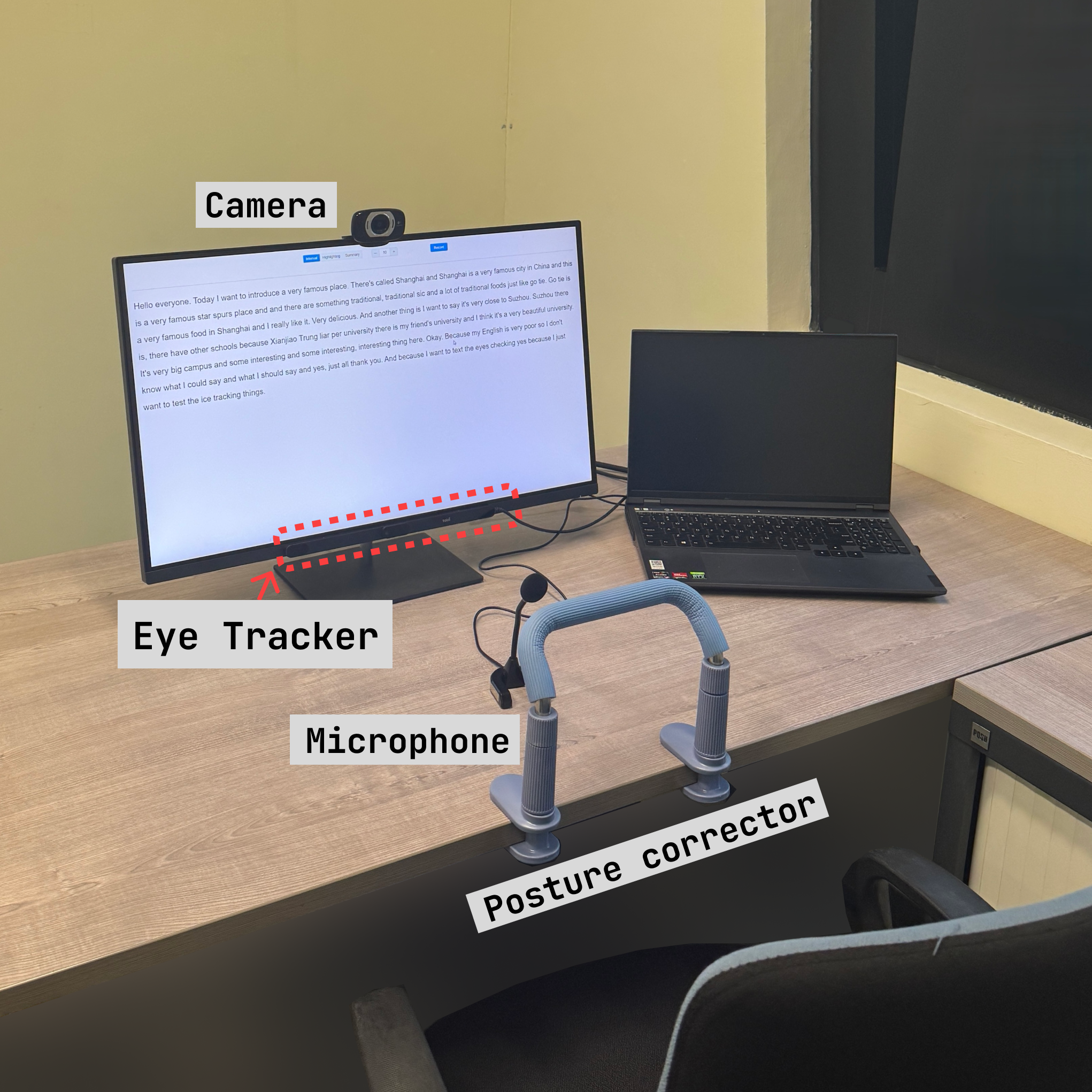}
  	\caption{Experiment Environment}
  	\label{fig:environment}
  \end{subfigure}%
  \hfill%
  \begin{subfigure}[b]{0.73\textwidth}
  	\centering
  	\includegraphics[width=\textwidth]{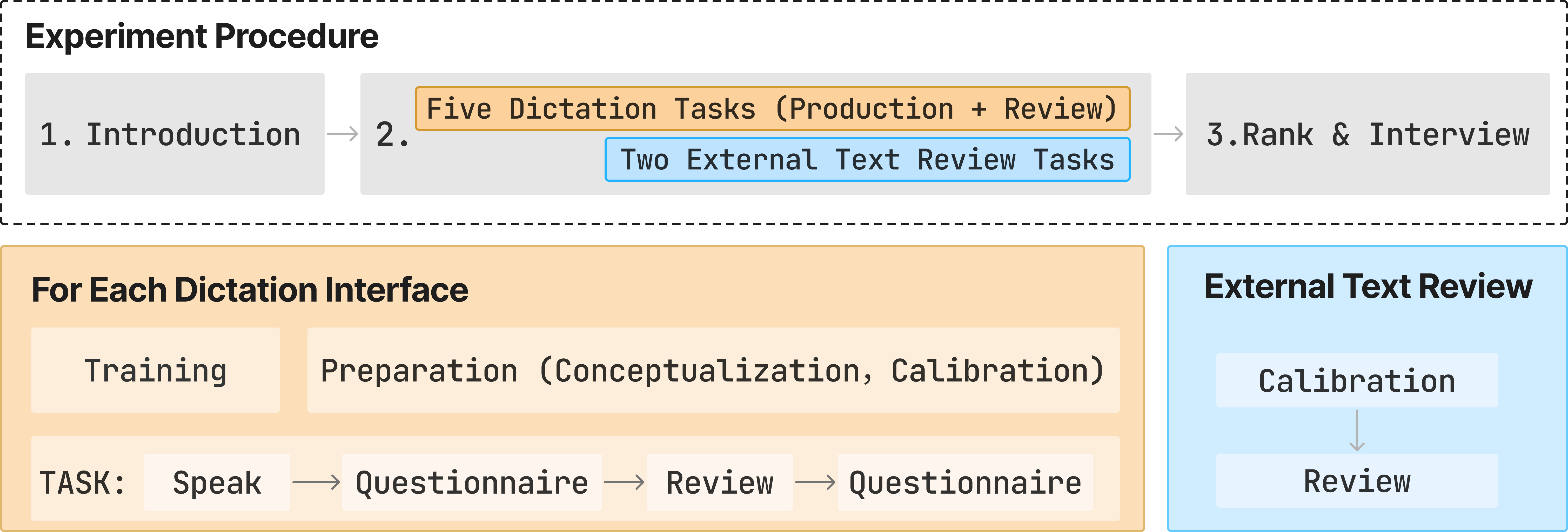}
  	\caption{Procedure}
  	\label{fig:procedure}
  \end{subfigure}
  
  \caption{Experiment environment and procedure. \toc{Figure (a) shows the experiment setup, featuring a participant seated with a posture corrector, a microphone on the desk, and a 24-inch monitor with a Tobii Pro Spark eye-tracker, capturing gaze data at 60 Hz. Figure (b) shows the procedure of experiment tasks. For each \interface{} there are five dictation tasks, each testing one \interface{}. Each dictation task begins with a training and preparation phase, then a data collection phase involving two steps: speak and review (reading). Two external review (reading) tasks of GPT-generated text are inserted between dictation tasks in counterbalanced order. 
  }}
  \label{fig:Exp Design}
\end{figure*}

\subsection{Task}


The participants were instructed to write a social media post with each dictation interface. Topic options were work-related (such as introducing recent project content, learning experience, or insights in a certain field, etc.) or life-related topics (such as introducing hobbies, how to relax on weekends, etc.). Participants chose one of these two topics, but the specific content can be freely conceived according to the participants' own preferences and not limited to our examples. Furthermore, we encourage them to compose content they were familiar with and that was relevant to their lives. This was to prevent the participants from spending a lot of time thinking about things they haven't done. The content of the five pieces cannot be exactly the same, and each piece must be a new composition. In regard to text length, we require participants to fill more than a third of the entire screen (about 250-300 words). If they finished a short composition, the experimenter asked participant to compose again.


We adopted a task design strategy similar to a previous eye movement study on writing \cite{torrance2016reading} by separating the composition phase and the reviewing phase. This allowed us to instruct the participants effectively, e.g., making sure they carefully reviewed the text, and to be able to clearly separate the data from speech production and reviewing processes, which are expected to be very different.  

In the reviewing phase, we added two example pieces of text generated by ChatGPT in the comparison, to contrast the reading behaviors of self-produced/familiar text with reading external text. We created two pieces of external text in similar topics of the participants' composition, varied in reading difficulty for each participant. The prompt used is as follows, [topic] was replaced according to the participant's actual topic \rr{(example generated content is detailed in Appendix \ref{easy_and_hard_example})}: 

\begin{itemize}
  \item \easy{}: Create a social media post about [topic] using simple and direct language. The target audience is the general public, so avoid using complex jargon and long sentences.

  \item \hard{}: Create a social media post about [topic] using complex sentence structures and sophisticated vocabulary. Include professional jargon and abstract concepts to attract professionals.
\end{itemize}

This also serves as a benchmark for checking the legibility of eye-tracking data based on existing knowledge from previous reading studies.

\subsection{Apparatus}

As shown in Figure~\ref{fig:environment}, the experiment took place in a quiet room furnished with a standard desk and chair. A microphone was placed in front of the participant on the desk. 
To ensure the quality of gaze data collection, participants were asked to sit behind a sitting posture corrector, which helped them to center their bodies while preventing them from facing down and keeping an optimal distance with the eye tracker (around 65cm).  
A 24-inch monitor (60Hz) with 1920 $\times$ 1080 resolution was used to display the dictation interface. 
A Tobii Pro Spark eye-tracking device was attached to the bottom of the monitor, capturing gaze data at 60 Hz. Tobii Pro Lab was used for calibration and screen recording. 
The experiment software is a web application running on the Google Chrome browser. 

\subsection{Participants}

Participants were recruited at a university campus via online advertisements and snowball sampling. 
Eligible participants were required to have normal or corrected vision, and be fluent in English speaking, with a minimum requirement of scoring the IELTS test over 6.5. 
Twenty participants ($N = 20$) were recruited, comprising eleven self-identified females and nine self-identified males. Their ages ranged from 19 to 30 years ( $mean=23.75$ ), and they had varied experience with speech-to-text technology: two participants used it daily, nine participants used it occasionally, seven participants only tried it a few times, and two never used it. The participants' English language proficiency levels varied as well. Six participants (P1 - P6) self-identified as native / bilingual English speakers. Ten participants (P8 - P17) passed IELTS test scores $ \geq 6.5 $, and three (P18 - P20) passed TOEFL test scores $  \geq 102$. One received both high school and university education in English-taught programs. 




\subsection{Procedure}
Participants began with a brief introduction from the experimenter. After signing the informed consent form, their demographic information was collected, including their self-identified gender, age, and experiences with Speech-to-Text systems. 
Each participant performed five tasks, each involving a speech production phase and a reviewing phase, using a different \interface{} (\plain, \aoc, \rake, \gptsm, and \as). 
The orders of the five conditions across participants was counterbalanced using Latin Square. Two additional external text-reviewing tasks (\easy{}, \hard{}) were inserted into the first and second half of the experiment between trials, with their orders being shuffled across participants. 

For each \interface{} condition, the following steps were executed. 1) The experimenter provided a detailed explanation for the interface used in this condition. Participants performed a training task using this interface to freely compose text on any topic until they were familiar with its function and felt comfortable using it. 2) Then they were given a moment to conceptualize their idea of composition. 3) Before starting to capture data, they calibrated their gaze tracking using the Tobii Pro Eye Tracker Manager. 4) When participants were ready, they could start speaking after clicking the Start Recording button. After finishing the composition, they stopped the recording and filled in a short survey rating their experience and recalling their gaze behavior. 5) Next they were asked to carefully read the text from their composition from start to end, displayed in the respective \interface{} condition. To make sure participants read carefully, they were told to rate the quality of the text afterwards. 6) Participants answered another short survey after reviewing.

A semi-structured interview was conducted at the end of the experiment, where participants gave their rankings of the interfaces and provided subjective feedback about their experience. Upon completion, each participant received a local supermarket coupon as compensation.  

\subsection{Data Collection}

We collected both quantitative and qualitative data from the experiment. In the speech production process, we captured screen recordings via Tobii Pro Lab, which provides a comprehensive gaze matrix for further analysis. In the reviewing process, we used the Eye-Movement in Programming Toolkit (EMTK)\footnote{\url{https://github.com/nalmadi/EMIP-Toolkit}} to automatically create areas of interest around each word. 

For subjective feedback, we collected surveys rating participants' preference, frustration, satisfaction and mental effort for each \interface{} condition, in the speech production process and reviewing process respectively. Semi-structured interview further elaborated their experience with open-ended questions. Interviews were recorded via audio. 







\section{Data Preparation and Analysis Methods}

This section outlines our methods for processing the eye movement data and developing quantitative metrics for both the \emph{speech production} and \emph{reviewing} phases. We measured how participants visually engaged with the displayed text during and after dictation, and quantified their reading effort when reviewing the text.

\begin{figure*}

    \centering

    \begin{subfigure}{0.45\textwidth}
        \centering
        \includegraphics[width=\textwidth]{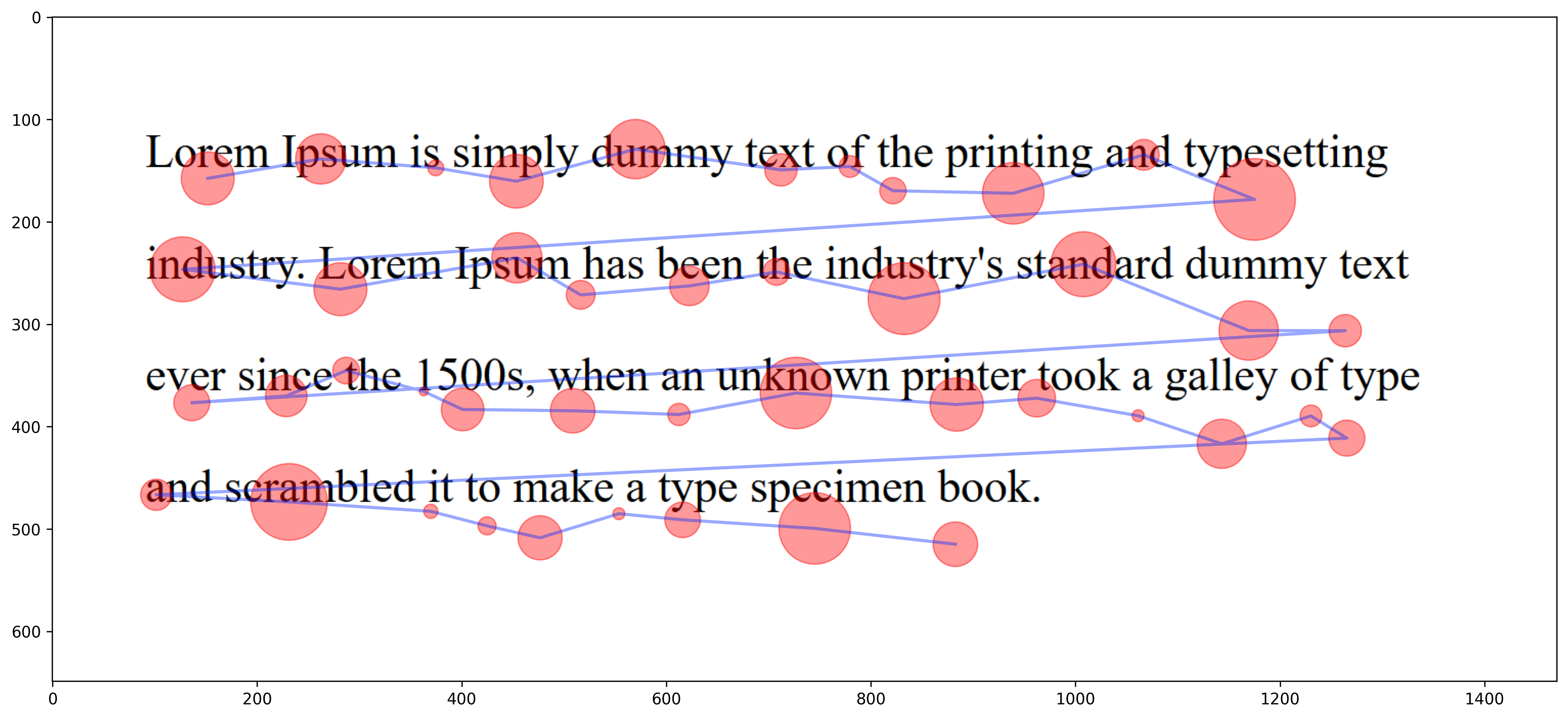}
        \caption{Eye movement data before drift correction.}
    \end{subfigure}
    \quad
    \begin{subfigure}{0.45\textwidth}
        \centering
        \includegraphics[width=\textwidth]{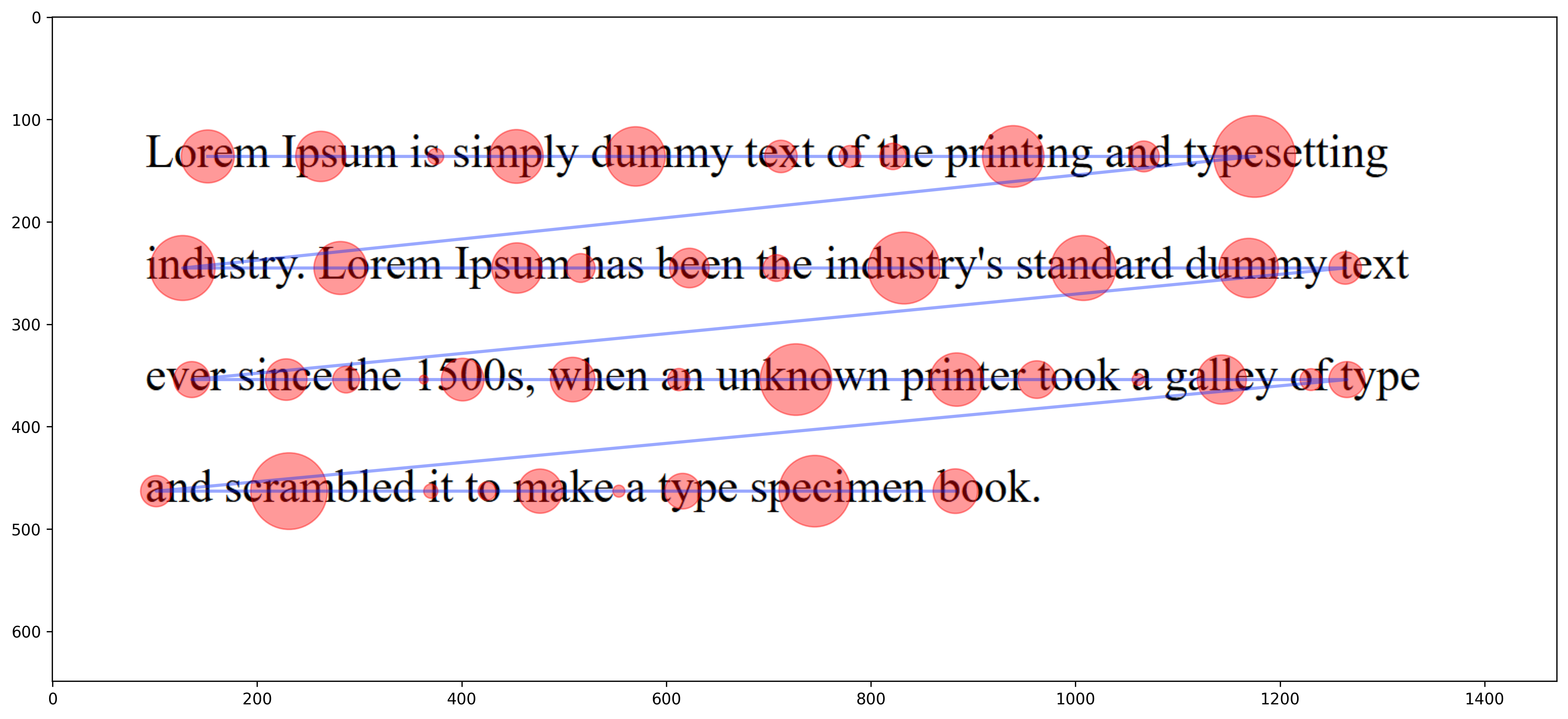}
        \caption{Eye movement data after drift correction.}
    \end{subfigure}

    \caption{Examples of drift correction where fixation positions are reattached to their original line of text. As shown in Figures (a) and (b), after drift correction, the drifted gaze points were adjusted, aligning the fixations to the text tokens. This is needed to accurately measure eye movement metrics over tokens of text.}
    \label{fig:drift_correction}
\end{figure*}

\subsection{Data Preparation}

\paragraph{\textbf{Annotation}} 

For the speech production phase, we manually annotated the TOI (Time of Interest) as the valid data timeline for analysis, with the start points of the TOI adjusted to skip periods where participants were still contemplating or communicating with the experimenter but had not yet begun speaking. Then we exported the raw data filtered by TOI from Tobii for data analysis. 

For the reviewing phase, we manually set the TOI. The start time is marked by the participant's first fixation on the first line of text, while the end time is determined by the last fixation appearing at the end of the last line of text. Moreover, Tobii automatically extracted AOI (Areas of Interest) to be individual words (with or without their adjacent punctuation). Each AOI is referred as a \emph{token} throughout this paper. We manually tagged tokens to differentiate between highlighted and non-highlighted words in RAKE and GP-TSM. Then, we exported the raw data with \emph{token}, tag of \emph{token}, and location of \emph{token} which were filtered by TOI from Tobii for data analysis.

\paragraph{\textbf{Cleaning and Correction}}
For the token-level data analysis, we used an open-source tool named Fix8\footnote{https://github.com/nalmadi/fix8/} to correct the gaze points in the raw output from Tobii, which greatly improved the alignment of gaze points and text tokens overcoming drift, as illustrated in Figure \ref{fig:drift_correction}.

\subsection{Eye Movement Metrics}

\paragraph{\textbf{Gaze Engagement Metrics}}

\begin{figure*}

    \centering

    \begin{subfigure}{0.45\textwidth}
        \centering
        \includegraphics[width=\textwidth]{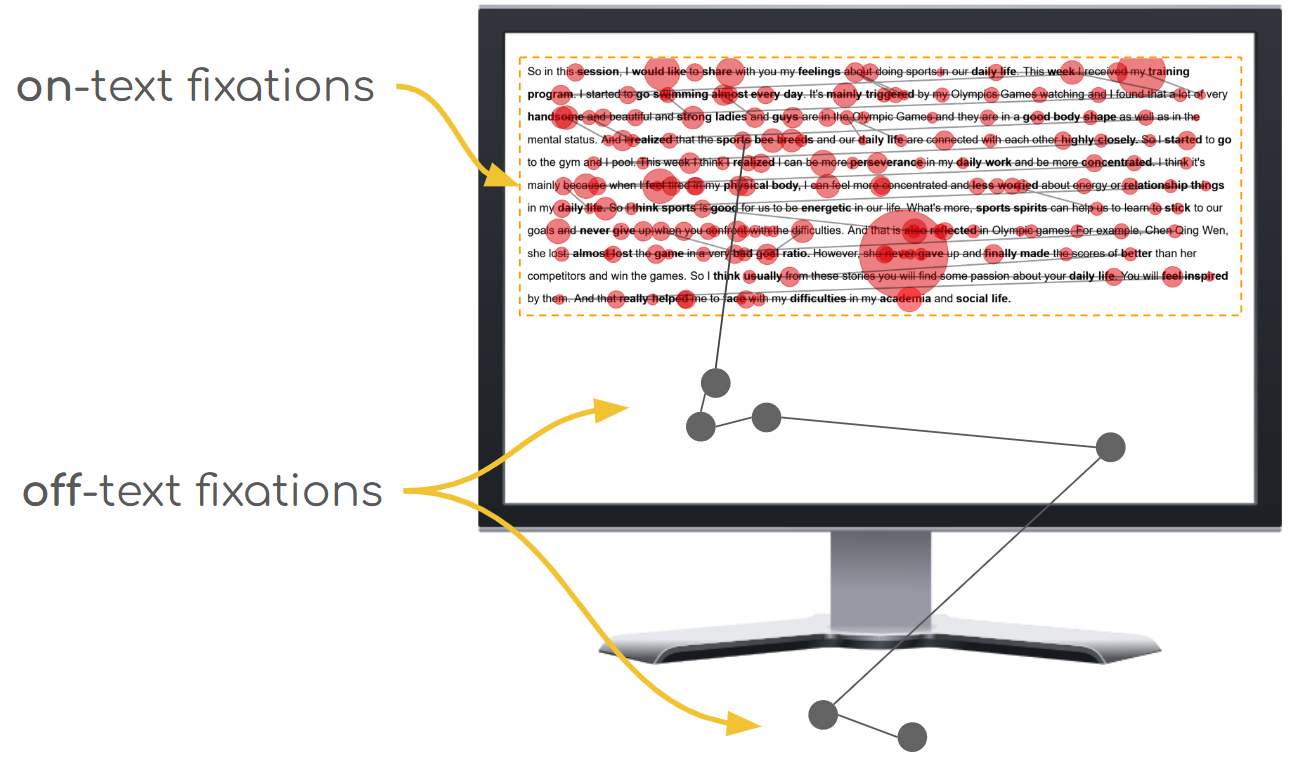}
        \caption{On-text in red and off-text fixations in gray.}
    \end{subfigure}
    \quad
    \begin{subfigure}{0.45\textwidth}
        \centering
        \includegraphics[width=\textwidth]{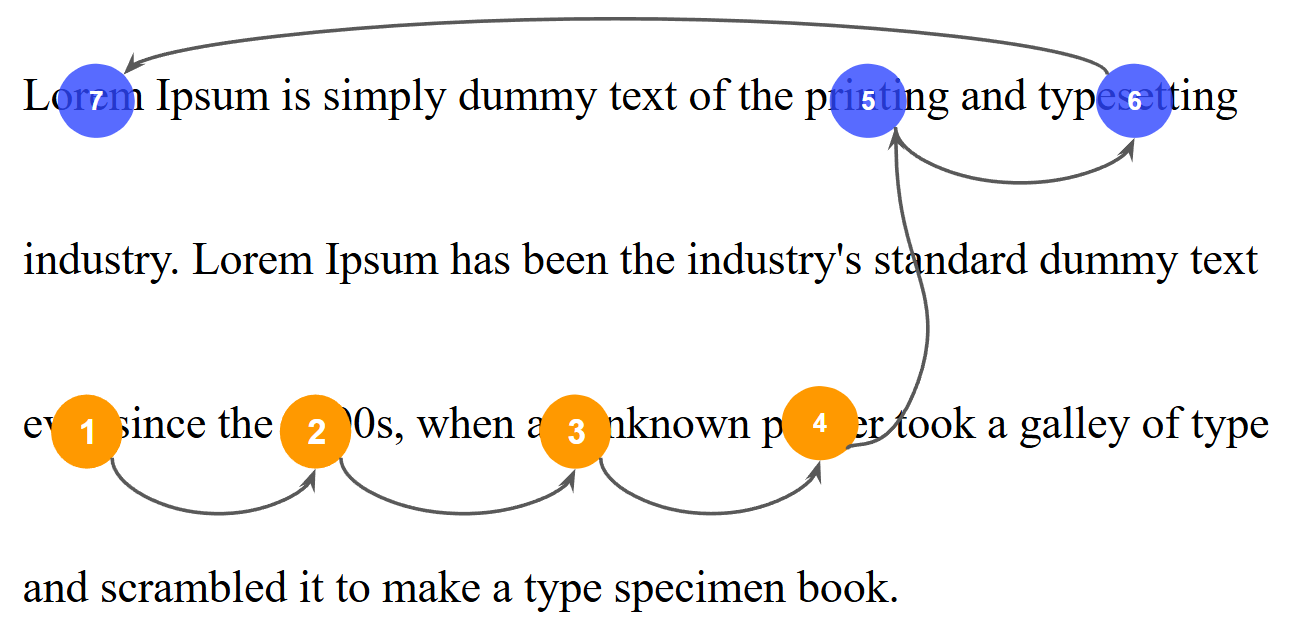}
        \caption{Sustained reading in orange and hopping in blue.}
        \label{fig:sustained_reading}
    \end{subfigure}

    \caption{\toc{Illustrating eye-movement engagement metrics, Figure (a) shows the distinction between fixations on-text and fixations off-text.  Figure (b) shows the distinction between sustained reading where fixations move from left to right in a typical linear reading order, and hopping where the eyes make jumps across the text.}}
    \label{fig:eye_metrics}
\end{figure*}

The distribution of the participants’ gaze points can be categorized into: 1) \offtext{}, where the user’s gaze is away from the text area of the system or even away from the monitor; 2) \ontext{}{}, when the user is looking at the text area. In this situation, it can be further divided into two categories: \sustained{} and \hopping{}. Sustained Reading describes fixation sequences that reflect typical reading patterns from left to right with short saccades between consecutive words \cite{torrance2016reading}. As seen in Figure \ref{fig:eye_metrics}~(b), sustained reading is present when sequences of three or more forward-moving (left to right) consecutive fixations exist with saccades that do not exceed 250 pixels.  This number of pixels was adopted from 25 letter spaces as described by \citet{torrance2016reading}. Hopping describes on-text eye movement that does not correspond with sustained reading, where the eyes ``hop'' from one word to another non-adjacent word without following a typical reading pattern (left to right) \cite{torrance2016reading}.

\paragraph{\textbf{Reading Metrics}}
To evaluate reading effort during the reviewing phase, we looked into two types of widely used metrics, the first is regressions (jumps back) (\regressionin{}, \regressionbt{}) at \emph{trial} level, and the second set of metrics are fixation duration and fixation count metrics at the \emph{token} level (individual words) (\ttt{}, \fc{}, \ffd{}). Relying on \cite{rayner1998eye}, we can define these metrics as follows:
\begin{itemize}
    \item \toc{\regressionin{}: When the eyes make a jump-back to a \textbf{previous word} on the same line.  The jump between fixation 6 and 7 in Figure \ref{fig:eye_metrics} (b) demonstrates a within-line regression.}

    \item \toc{\regressionbt{}: When the eyes make a jump-back to a previous word on a \textbf{previous line} other than the current line. The jump between fixation 4 and 5 in Figure \ref{fig:eye_metrics} (b) demonstrates a between-line regression.}

    \item \ttt{} (TT): The sum of all fixation durations on a word, including fixations from regressions (in milliseconds).
    \item \fc{} (FC): The total number of fixations on a word.
    \item \ffd{} (FFD): The duration of the first fixation on a word (in milliseconds).
    
\end{itemize}

\subsection{\textbf{Statistical Analysis Methods}}

We conducted normality tests for all the measures used in this experiment, by first using the Shapiro-Wilk test (p > 0.05), and then using Levene's test (p > 0.05) to assess homogeneity of variances. Our choice of statistical analysis method for a measure depended on whether normality is violated. \rr{Information of normality tests and selected statistical analysis method were detailed in Appendix C table \ref{tab:Normality_test}.}
\rr{We also tested potential ordering effect caused by fatigue or boredom and found no significant effect across all trial-level measurements (details in Table \ref{tab:ordering_effect}).}

For multiple comparisons, ANOVA was used for normally distributed data with equal variances (p > 0.05), with Eta Squared estimating the effect size. If ANOVA results were significant (p < 0.05), post-hoc analysis was performed using Tukey HSD. For measures violating normality, the Kruskal-Wallis test was applied, using Epsilon Squared to estimate effect size, followed by Dunn's test for post-hoc comparisons if the results were significant. For pairwise comparisons, independent t-tests were applied when normality and equal variances were confirmed, with rank-biserial correlation (r) as the effect size. For non-normal data or unequal variances, the Mann-Whitney U test was used, with rank-biserial correlation (r) as the effect size. We used Bonferroni Correction to adjust for multiple tests.

\begin{figure*}
\centering

\begin{subfigure}[b]{0.32\textwidth}
   \centering
   \includegraphics[width=\textwidth]{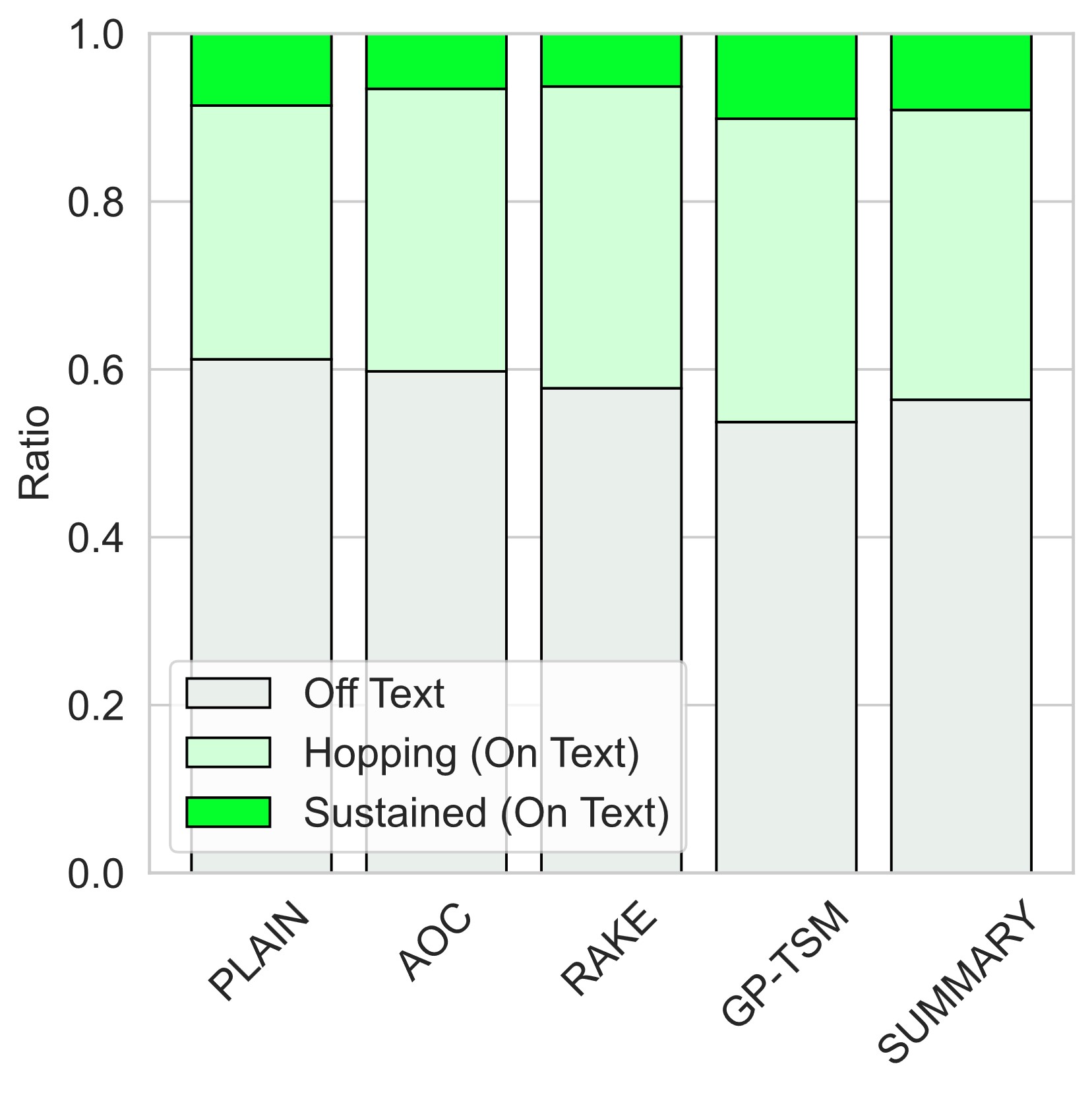}
   \caption{Gaze points ratios in speech production.}
   \label{fig:speak_off_text_ratio}
\end{subfigure}%
\begin{subfigure}[b]{0.32\textwidth}
   \centering
   \includegraphics[width=\textwidth]{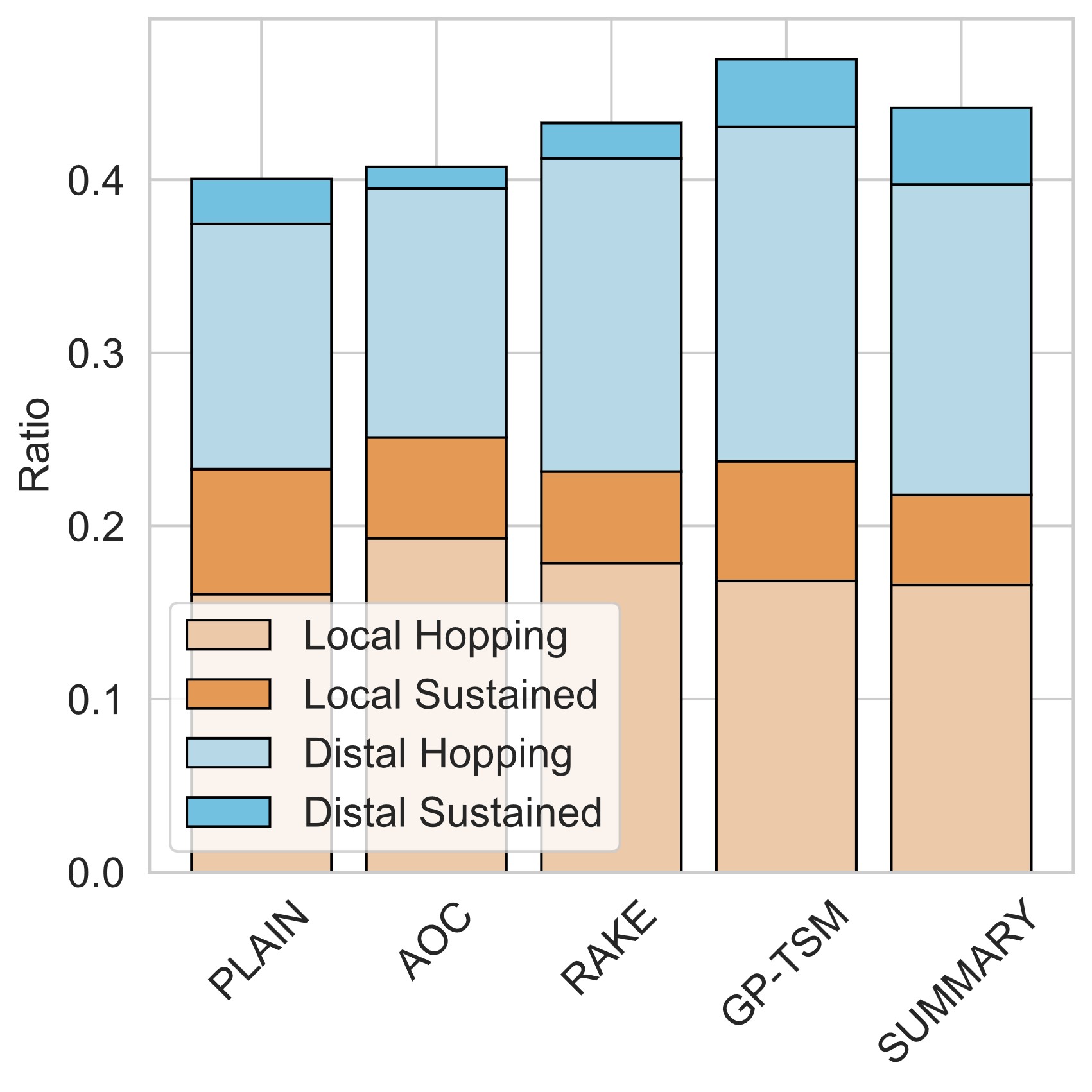}
   \caption{Distal and local gaze in production.}
   \label{fig:speak_distal_local_ratio}
\end{subfigure}%
\begin{subfigure}[b]{0.32\textwidth}
   \centering
   \includegraphics[width=\textwidth]{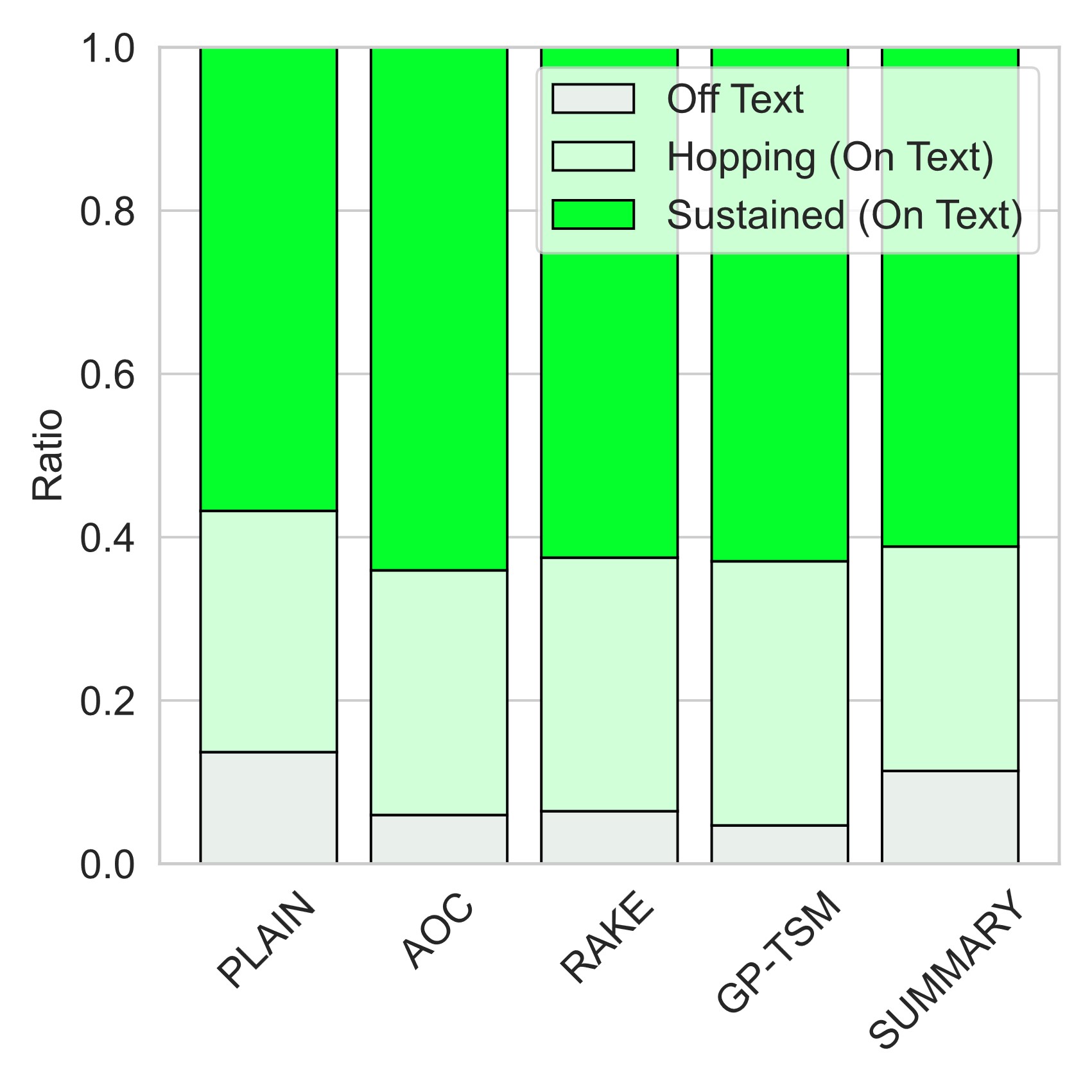}
   \caption{Gaze points ratios in reviewing.}
   \label{fig:review_off_text_ratio}
\end{subfigure}

\caption{Gaze Engagement during speech production and review across five \interface{} conditions. a): Average percentages of gaze points \ontext{} (in grey) and \offtext{} (in green) during speech production. Two shades of green show percentage of \sustained{} and \hopping{}. b): Detailed Breakdown of \ontext{}{} Gaze Points distinguishing between \local{} and \distal{} from the position of production (cursor position). c) Average percentages of gaze points \ontext{} (in grey) and \offtext{} (in green) during reviewing.}
\label{fig:gaze_engagement}
\end{figure*}

\subsection{\rr{Qualitative Analysis Method}}
\rr{We analyzed participants' interview recordings using Reflexive Thematic Analysis \cite{braun2019reflecting}. The interview audio was transcribed using a speech-to-text system and manually verified for accuracy. We employed a hybrid deductive-inductive approach for analysis. Initially, feedback was categorized deductively based on the distinct task conditions and phases. Subsequently, two authors independently conducted open coding on the transcripts to inductively identify emerging insights and patterns. Discrepancies in coding were resolved through iteration. Finally, these codes were iteratively grouped into themes as our interpretations of user feedback.} 

\section{Eye Movement Results}
This section reports the findings from the analysis of eye movement in terms of gaze engagement patterns and reading efforts. Gaze engagement patterns are observed from whether gaze points are on or off text, whether they are sustained reading or hopping as well as their locality. Reading efforts are measured by fixation time and counts.

\subsection{\rr{RQ1: How do participants visually engage with the text during speech production versus reviewing?}}
\label{chap:eq1_users_visually_engage}

\paragraph{\textbf{Sustained Reading vs. Hopping.}} Figure~\ref{fig:speak_off_text_ratio} illustrates the gaze behavior observed during speech production. We can observe that overall participants spent less than 56\% of the production time \ontext{}{}, with a range from 39\% to 56\% across \interface{}. \sustained{} only took up 7\% to 11\% of the production time, while the rest of the \ontext{} time was spent on \hopping{} (30\%-36\% of production). 
To observe the differences between speech production and reviewing processes in gaze engagement, we also calculated for the reviewing process the ratios of \ontext{} (range 86\%-95\%), \sustained{} (range 56\%-64\%) and \hopping{} (range 27\% to 32\%), as illustrated in Fig. \ref{fig:review_off_text_ratio}.

\paragraph{\textbf{Speaking vs. Reviewing.}} 
Mann-Whitney U tests on \offtext{} were performed to compare between \emph{speaking} and \emph{reviewing}. It indicated that \emph{speaking} \meanse{0.578}{0.013} had a 5.9 times 
higher ratio than \emph{reviewing} \meanse{0.084}{0.017} \mannwhitneyu{13582.5}{12.4}{0.801}.
Mann-Whitney U tests on \sustained{} were performed to compare between \emph{speaking} and \emph{reviewing}. It indicated that \emph{speaking} \meanse{0.089}{0.007} had a 86.71\% lower \sustained{} ratio than \emph{reviewing} \meanse{0.668}{0.004} \mannwhitneyu{0}{-13.2}{-0.85}. 
The same comparisons between \emph{speaking} and \emph{reviewing} on \offtext{} and \sustained{} were conducted for each individual \interface{}. The findings follow the same trends as for all interfaces combined, details can be found in Appendix C Table~\ref{tab:speak_vs_review}.

Moreover, Mann-Whitney U tests on \hopping{} were performed to compare between \emph{speaking} and \emph{reviewing}. Interestingly we found no significant difference between them (p=0.266). This also remains the same for each \interface{}, specifically \plain{}: p=0.598, \aoc{}: p=0.818, \rake{}: p=0.715, \gptsm{}: p=0.818, \as{}: p=0.663. The average \hopping{} ratio during speaking \meanse{0.341}{0.011}, close to that during reviewing \meanse{0.332}{0.004}. 


\paragraph{\textbf{Distal vs. Local in speaking.}} Figure~\ref{fig:speak_distal_local_ratio} dives further into the \ontext{} gaze behaviors by dividing them into \local{} and \distal{} situations, calculated with the distance between the gaze position and the position of the line of text being produced. A fixation or a sequence of fixations is considered \local{} if its position starts on the line of text currently being produced, and is considered \distal{} if the fixation or sequence starts on a previous line other than the line currently being produced. This distinction was adopted from a previous study on eye movement during written text production \cite{torrance2016reading}. Therefore we could observe the ratios of \distal{} ranging from 16\% to 23\%, within which  \distal{} \sustained{} ranging from 1\% to 4\%, \distal{} \hopping{} ranging from 14\% to 19\%, \local{} \sustained{} ranging from 5\% to 7\%, and \local{} \hopping{} ranging from 16\% to 19\%.

\paragraph{\textbf{Answers to RQ1.}} This section described the gaze distribution on and off the text, with a distinction between sustained reading and hopping as well as its locality. 

\toc{During production, participants spent less than 39-56\% of the time looking at text, within which only 7-11\% was sustained reading. Probably during speech production, much of the participants' cognitive bandwidth was occupied by composition, leaving little space for reading in parallel. It is known that high cognitive load is involved in retrieving content from memory, formulating intentions, and translating linguistic representation during production \cite{beers2010adolescent}. We also found that participants spent about half of their gaze-on-text time following the cursor (local gaze). 
When not following the cursor, their on-text gazes were mostly hopping around instead of performing sustained reading. While reading around the cursor allows for error-detection processes and facilitates word-level planning, distal reading ``may allow writers to compose texts with greater overall coherence in accordance with their rhetorical goals'' \cite{beers2010adolescent}. }

%


While reviewing, participants likely focus on reading the text thoroughly for comprehension, leading to 68\% sustained reading. This thorough reading is different from the short bursts of sustained reading performed during production for monitoring, error correction, and planning the next sentence.



\subsection{RQ2: How do the interfaces affect participants' visual engagement with text?}


In speech production, we performed Kruskal-Wallis tests comparing across five \interface{} conditions, for all the ratios including \ontext{} (p=0.883), \sustained{} (p=0.938) and \hopping{} (p=0.479) across \interface{} and found no significant difference. Kruskal-Wallis tests for the ratios of \distal{} \sustained{} (p=0.197), \distal{} \hopping{} (p=0.555), \local{} \sustained{} (p=0.332), and \local{} \hopping{} (p=0.479) also found no significant difference across five \interface{}. 
This said, we found no significant effect of \interface{} on their gaze engagement patterns with the text while speaking. 
In reviewing, we conducted Kruskal-Wallis tests comparing across five \interface{} conditions, for the ratios of \ontext{} (p=0.809) and \sustained{} (p=0.572). As normality was not violated, a one-way ANOVA test was performed for \hopping{} (p=0.164) comparing across \interface{}, with no significant effect found.
Therefore, we also found no effect of \interface{} on their gaze engagement patterns with the text while reviewing.

\paragraph{\textbf{Answers to RQ2.}}
The above comparisons revealed that the interfaces we tested did not significantly affect how participants visually engaged with the text areas overall between on and off text, and between reading and hopping \toc{as well as their locality. This indicates the persistence of gaze strategies regardless of how spoken text is displayed.} 
We found this result surprising. First, we expected \aoc{} would help participants read with its corrected transcript. But \aoc{} had even slightly less \sustained{} than \plain{} (not significant) during speaking, and slightly more (also not significant) during review. We also expected highlights (\rake{} and/or \gptsm{}) would attract participants' attention to look back at the previous production more. Although we could see some trends in slightly higher \distal{} gaze ratios of these two conditions than others, no significant difference was found either. Therefore we conclude, our tested differences in the textual display, regardless of their fidelity to original speech, aid for review or interface responsiveness and stability, do not change how participants allocate their attention and effort between reading and other gazing activities.

\begin{figure*}
    \centering
\begin{subfigure}[b]{0.40\textwidth}
   \centering
   \includegraphics[width=\textwidth]{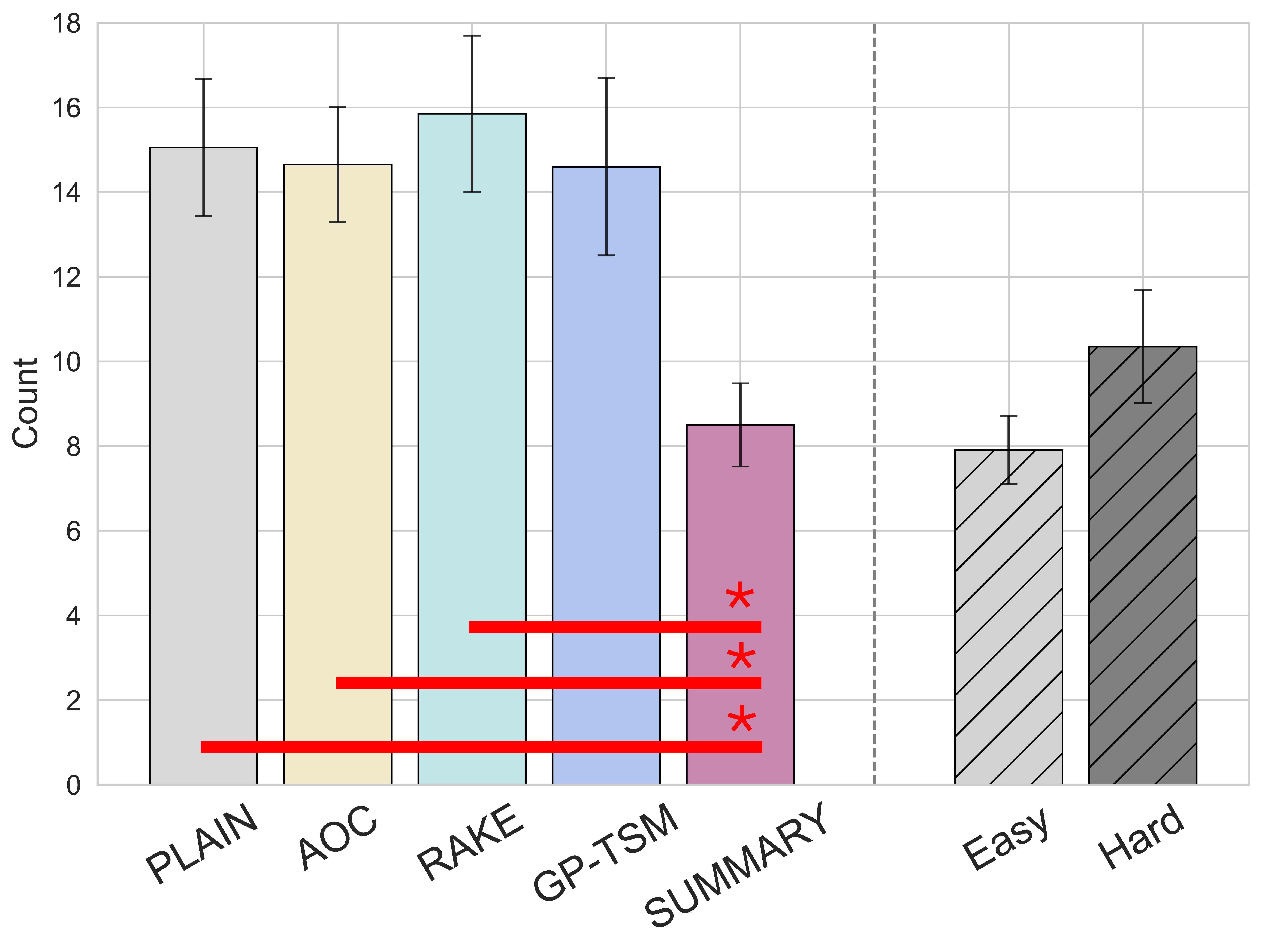}
   \caption{\regressionin{}}
   \label{fig:regression_within}
\end{subfigure}%
\quad
\begin{subfigure}[b]{0.40\textwidth}
   \centering
   \includegraphics[width=\textwidth]{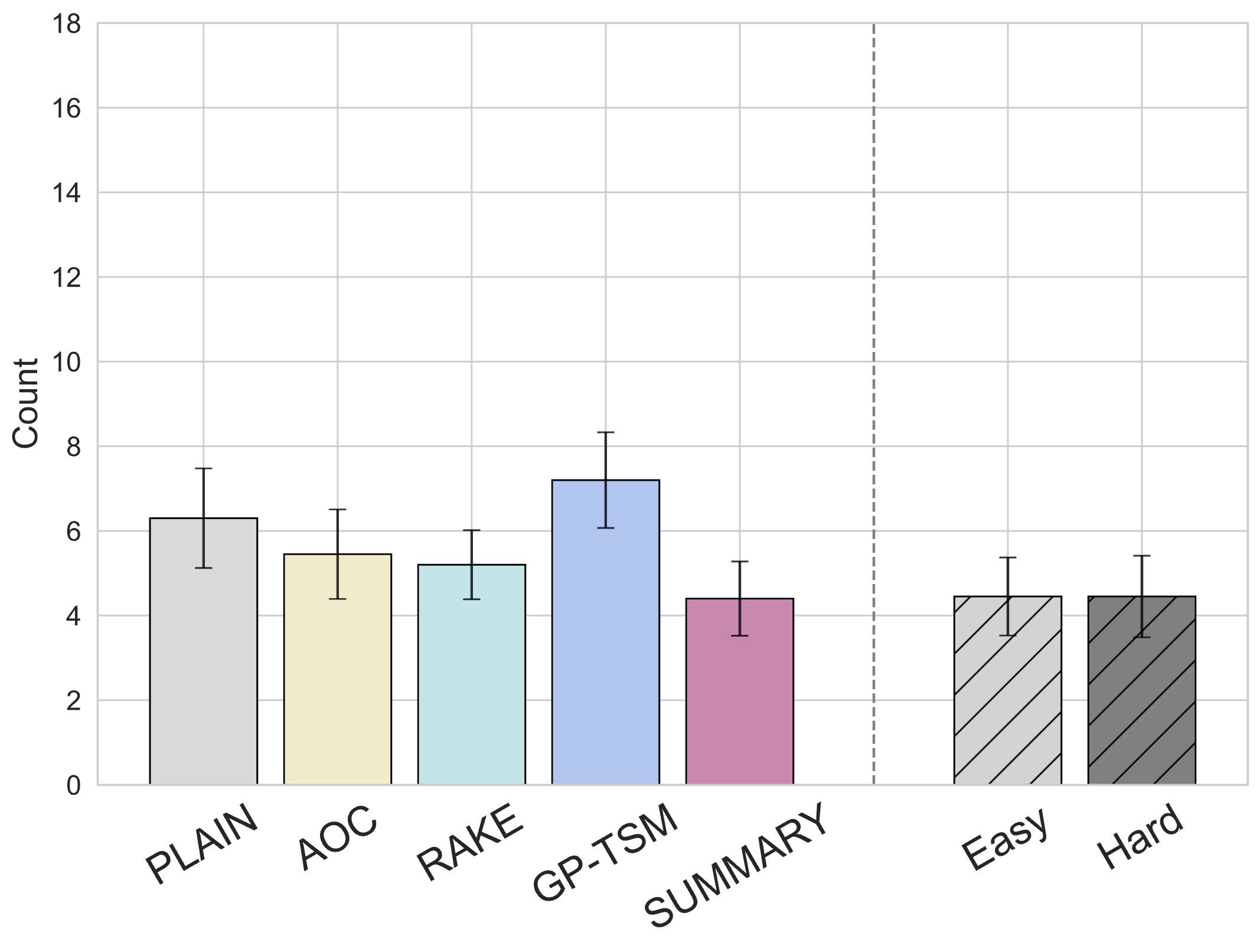}
   \caption{\regressionbt{}}
   \label{fig:regression_between}
\end{subfigure}%
        \caption{\toc{Regression metrics across five \interface{} conditions, with two examples of external text for reference. For \regressionin{} in Figure (a), \as{} is significantly lower and \plain{}, \aoc{} and \rake{}. For \regressionbt{}{} in Figure (b), we did not find any statistically significant difference across \interface{}. Both figures have the regression measures for reading GPT-generated external text (\easy{} and \hard{}) for references.}} 
        \label{fig:regression_within}
\end{figure*}

\begin{figure*}
\centering

\begin{subfigure}[b]{0.32\textwidth}
   \centering
   \includegraphics[width=\textwidth]{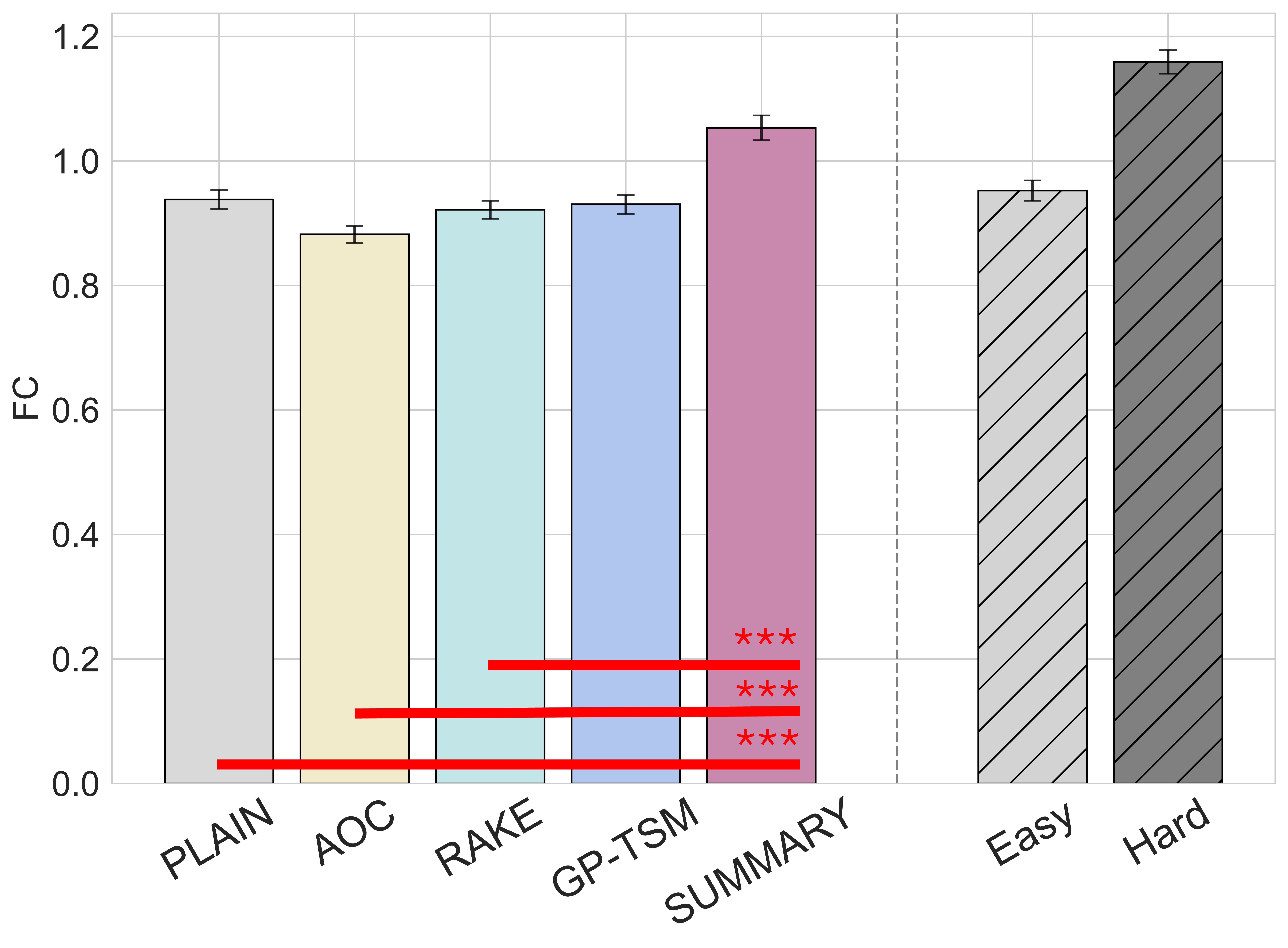}
   \caption{Fixation Count (FC).}
   \label{fig:all_fc}
\end{subfigure}%
\begin{subfigure}[b]{0.32\textwidth}
   \centering
   \includegraphics[width=\textwidth]{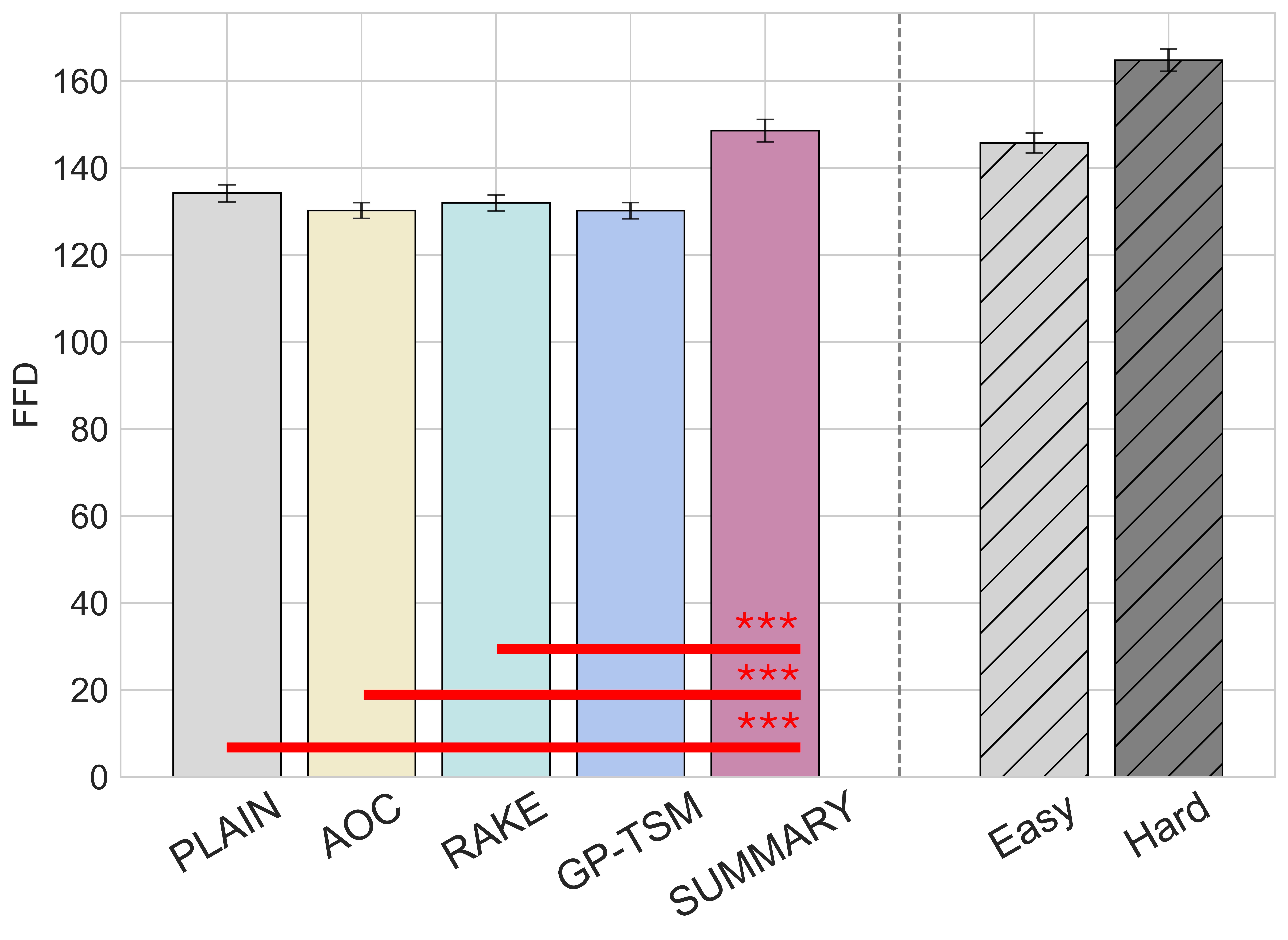}
   \caption{First-Fixation Duration (FFD).}
   \label{fig:all_ffd}
\end{subfigure}%
\begin{subfigure}[b]{0.32\textwidth}
   \centering
   \includegraphics[width=\textwidth]{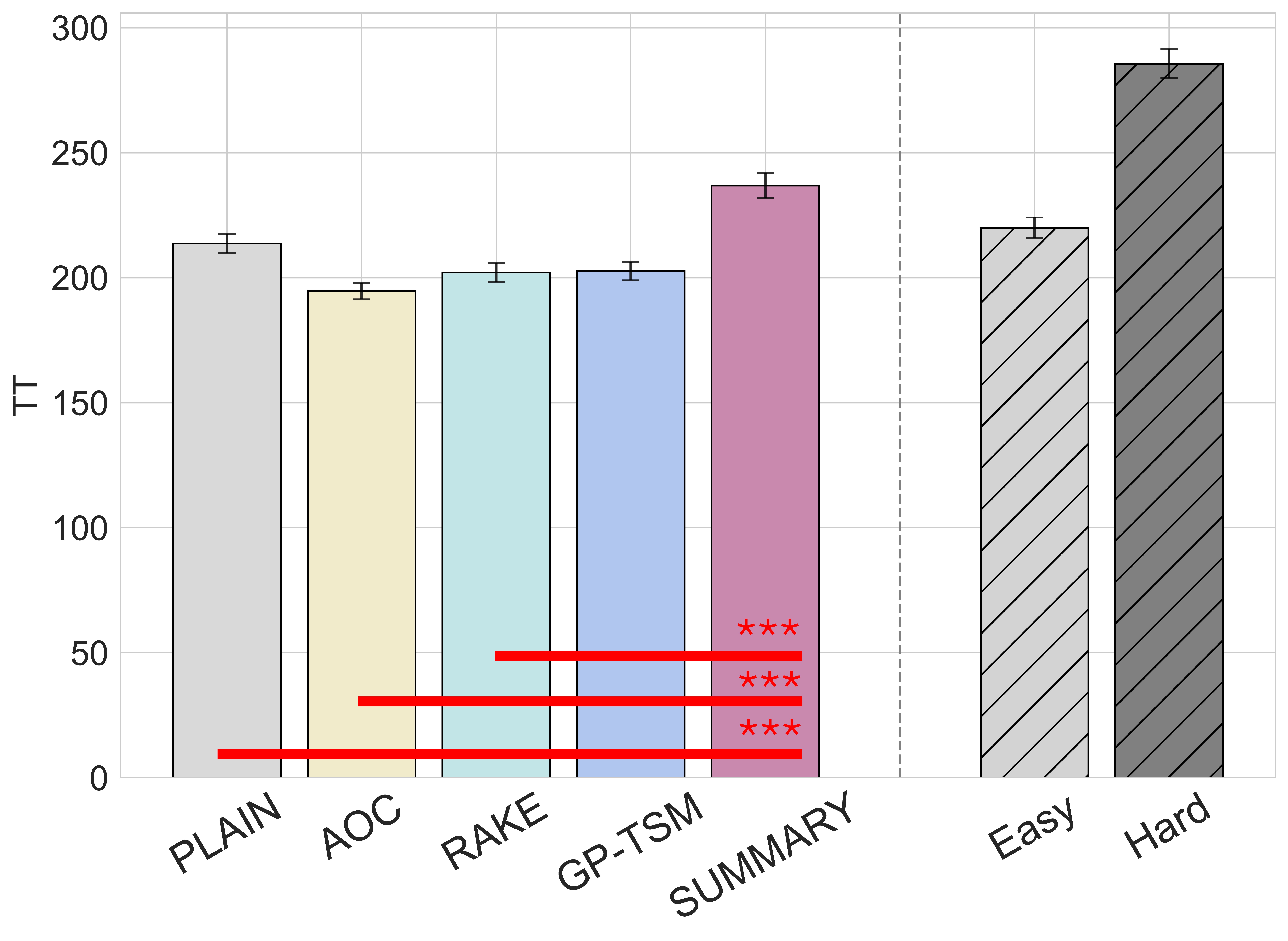}
   \caption{Total Time (TT).}
   \label{fig:all_tt}
\end{subfigure}
\caption{Eye Movement metrics results across five \interface{}, with two external text examples shown as references (duration in milliseconds). \toc{Statistically significant differences were found between \as{} and \plain{}, \aoc{} and \rake{} on FC, FFD and TT.}}
\label{fig:all_token_review}
\end{figure*}

\subsection{\rr{RQ3: How do the interfaces affect reading effort and patterns during reviewing?}}

\paragraph{\rr{\textbf{Regression.}}}
We compared both \regressionin{} and \regressionbt{}, where \regressionbt{} was normalized by the number of lines. Only \regressionin{} showed a statistically significant difference. 
Figure~\ref{fig:regression_within} reports the results on \regressionin{} across \interface{} conditions, with two external text examples as additional baselines for comparison. We performed a Kruskal-Wallis test for \regressionin{} across \interface{} and found a significant effect \kruskalnse{6}{30.52}{.023}{.18}\footnote{0.01 is considered a small effect, 0.06 medium and 0.14 large \cite{lakens2013calculating}.}. 
Dunn’s post-hoc test identified significant differences between \as{} and \plain{} (p=0.048), \aoc{} (p=0.032), \rake{} (p=0.030). Similarly, we also found significant differences between \easy{} and \plain{} (p=0.009), \aoc{} (p=0.006), \rake{} (p=0.005). \as{} has a lower average \regressionin{} \meanse{8.50}{0.98} than \plain{} \meanse{15.05}{1.61}, \aoc{} \meanse{14.65}{1.36}, \rake{} \meanse{15.85}{1.85}.


\paragraph{\rr{\textbf{Fixation metrics.}}} Figure~\ref{fig:all_token_review} reports the results on fixation metrics across \interface{} conditions, with two external text examples as additional baselines for comparison. We conducted Kruskal-Wallis tests across five \interface{} conditions, and found significant differences for all three fixation metrics. Specifically, 
    FC: \kruskalsse{4}{60.15}{}{.002}, 
    FFD: \kruskalsse{4}{42.44}{}{.001},  
    TT: \kruskalsse{4}{64.62}{}{.001}. 
Dunn’s post-hoc test found significant differences between \as{} and all three other interfaces (\plain{}, \aoc{}, \rake{}) on all the metrics, all the P values are p<0.001. We can see that \as{} has higher average values in FC, FFD and TT than the other interfaces, with differences ranging from 13\% to 16\%. 
Refer to Appendix C Table \ref{tab:token_level_metrics} for detailed values. This shows that \as{} is ``harder'' to read in terms of fixation measures, which indicates unfamiliarity.




\rr{\textit{For Native English Speakers.}} To explore whether language proficiency might influence our findings, we performed the same statistical analysis on a subset of the data from the six native English-speaking participants as detailed in Appendix \ref{appendix_native}. 
Some of the significant effects remained: \as{} had significantly higher FC than \aoc{} and \rake{}, and higher TT than \aoc{}. This confirms the finding about \as{} introducing unfamiliar words for this group of participants.
A new finding here is that \plain{} also had significantly higher FC and TT than \aoc{} and \rake{}, whereas these were not significant for the data of all participants. While more data is needed to draw conclusions on language proficiency, our result indicates a possibility that issues in raw transcripts could have a larger impact on the reading experience for native speakers.

\begin{figure*}
\centering

\begin{subfigure}[b]{0.32\textwidth}
   \centering
   \includegraphics[width=\textwidth]{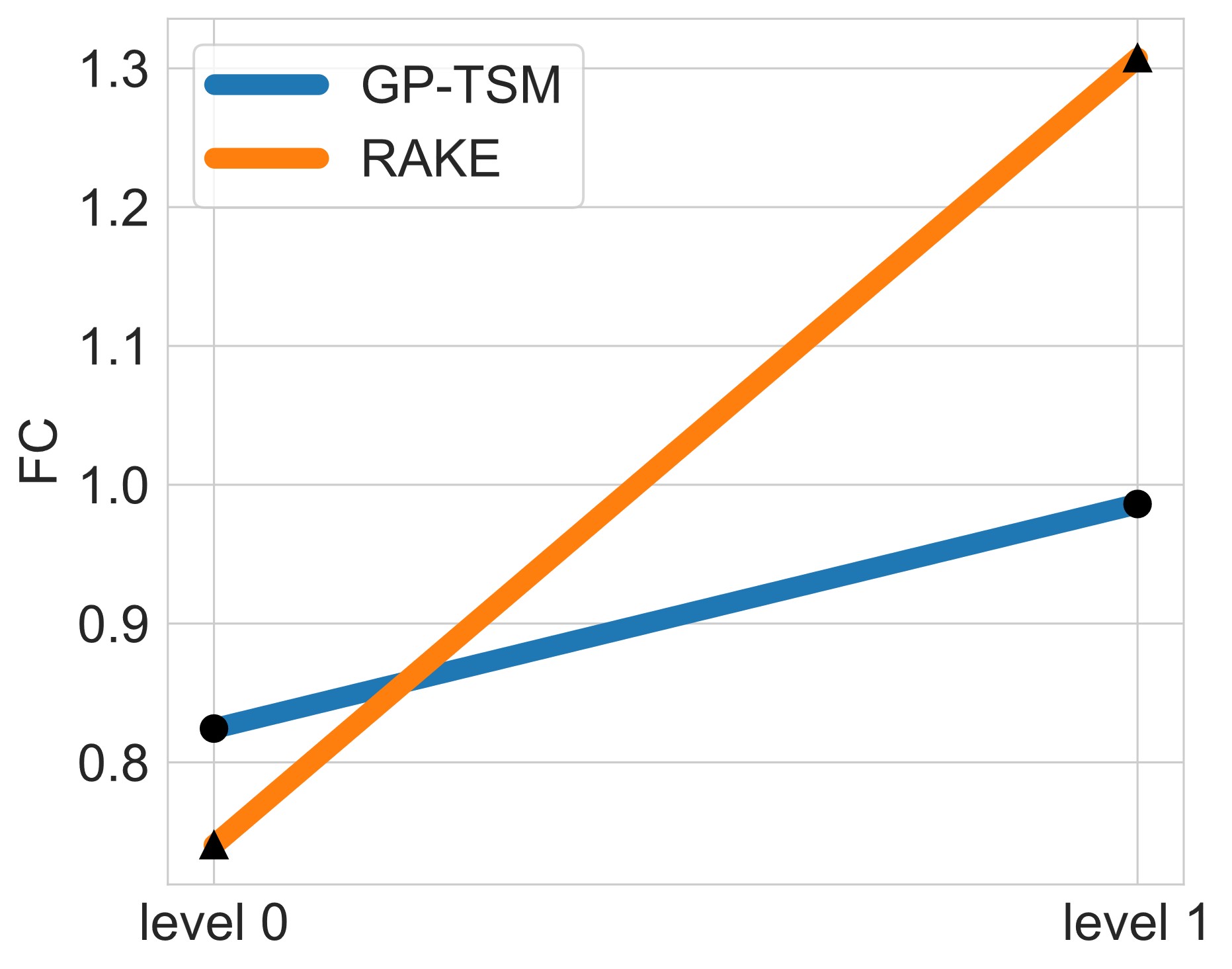}
   \caption{Fixation Count (FC).}
   \label{fig:fc_line}
\end{subfigure}%
\begin{subfigure}[b]{0.32\textwidth}
   \centering
   \includegraphics[width=\textwidth]{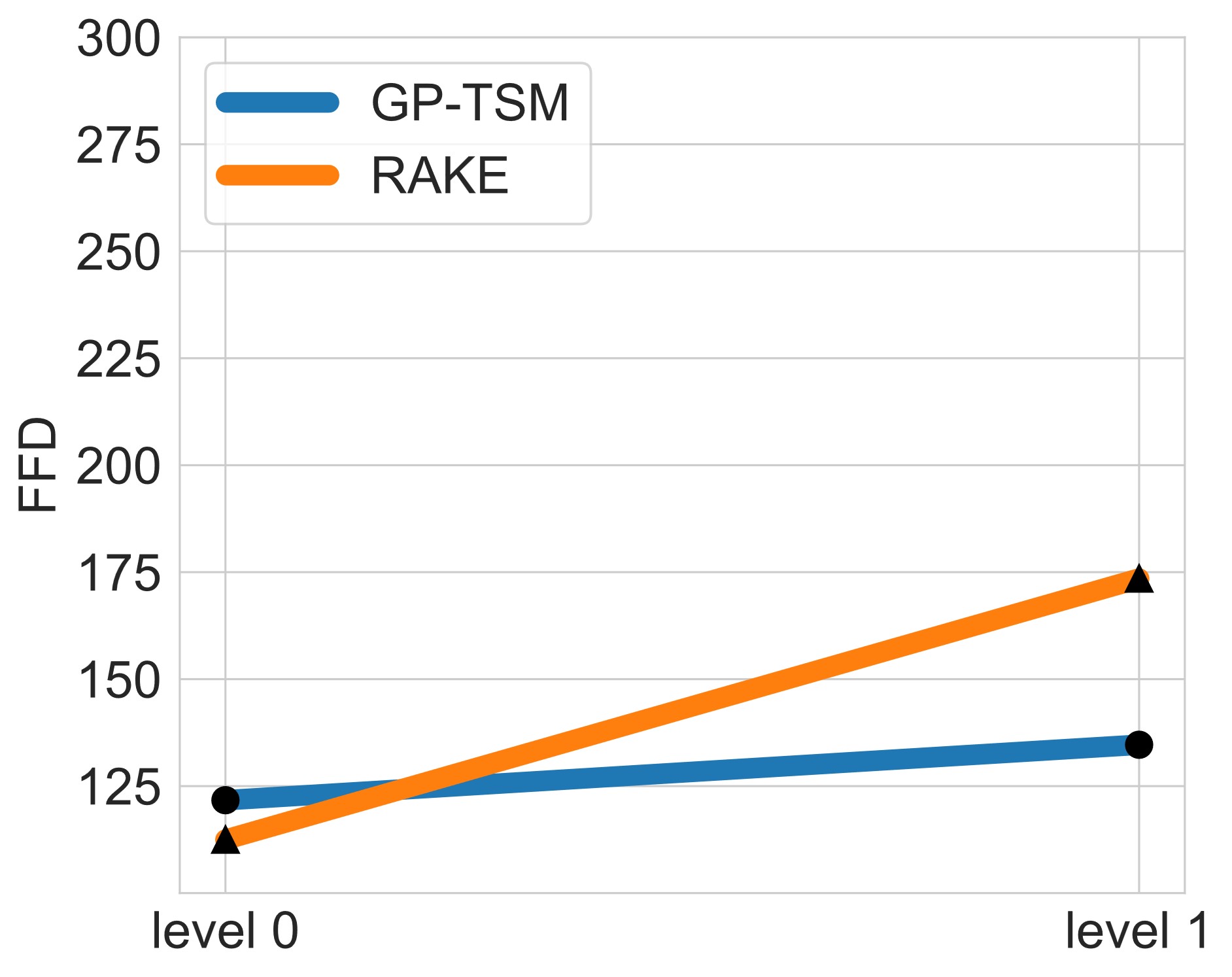}
   \caption{First-Fixation Duration (FFD).}
   \label{fig:ffd_line}
\end{subfigure}%
\begin{subfigure}[b]{0.32\textwidth}
   \centering
   \includegraphics[width=\textwidth]{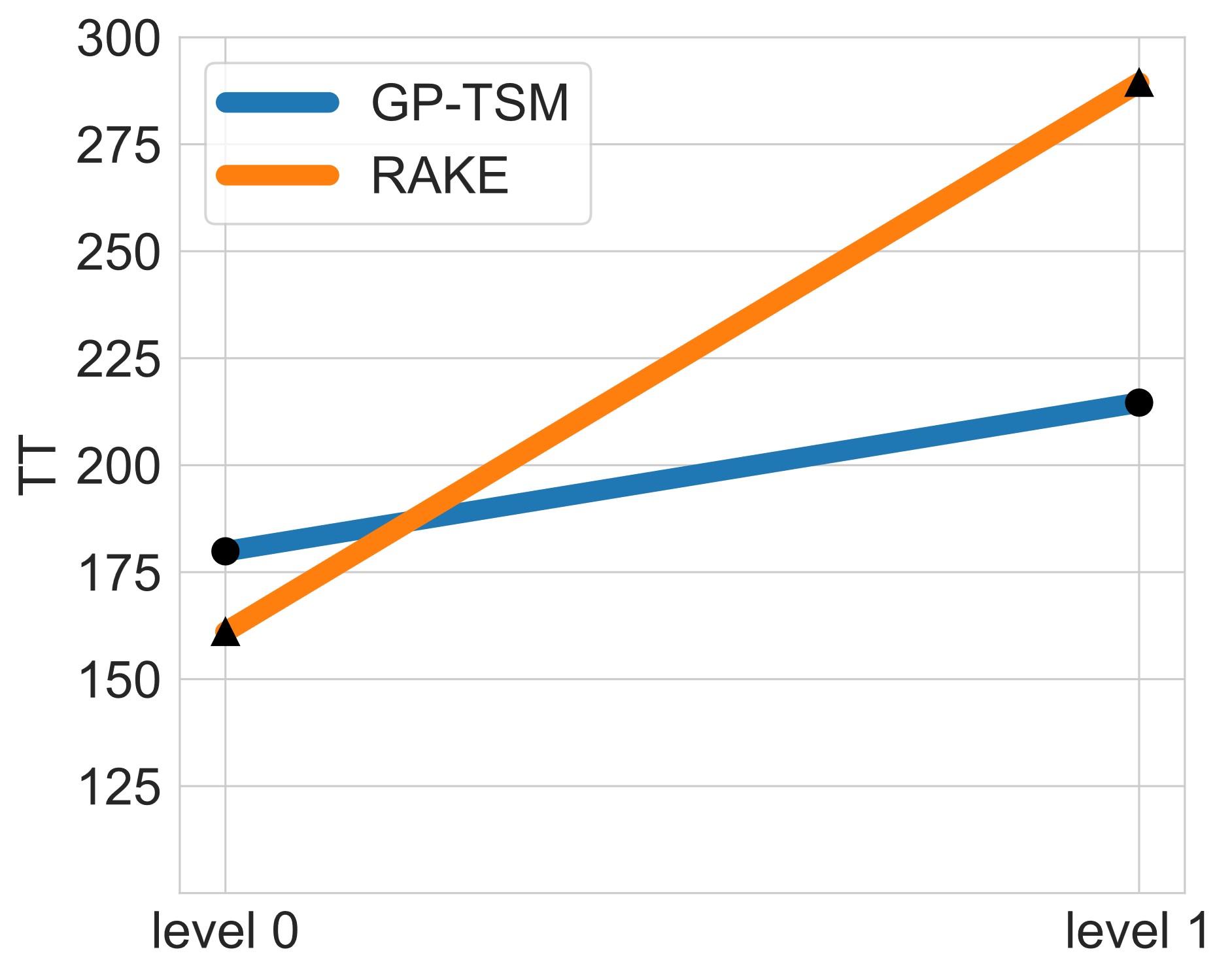}
   \caption{Total Time (TT).} 
   \label{fig:tt_line}
\end{subfigure}

\caption{FC, FFD (milliseconds) and TT (milliseconds) for \rake{} and \gptsm{} between highlighted (Level 1) and not highlighted (Level 0). \toc{For every measure, there is a significant interaction effect between \highlight{} and \level{}. 
\rake{} highlighting led to a greater increase from \zero{} to \one{} in fixation measures compared to \gptsm{}, suggesting \rake{} guides reading more effectively.}} 

\label{fig:highlight_guidance}
\end{figure*}

\paragraph{\rr{\textbf{RAKE vs. GP-TSM}}}
We would like to understand which highlighting technique could better guide participants' reading. Here we distinguish highlighted words from not highlighted words by calling them \one{} and \zero{}.
We applied Box-Cox transformation for FC, FFD and TT for \rake{} and \gptsm{} to correct their normality. Then we conducted a two-way ANOVA to investigate the main effects of \highlight{} method (\rake{}, \gptsm{}) and \level{} (\zero{}, \one{}), as well as their interaction effect on fixation.
The main effect of \highlight{} was not significant across all metrics (FC: p=0.272, FFD: p=0.140, TT: p=0.259), which indicates that \rake{} and \gptsm{} are not significantly different.
The main effect of \level{} was significant across all metrics (FC: \anovazh{1.0}{76.0}{20.737}{ < 0.001}{0.214}, FFD: \anovazh{1.0}{76.0}{19.312}{ < 0.001}{0.203}, TT: \anovazh{1.0}{76.0}{19.597}{ < 0.001}{0.205}). 
The interaction effect between \highlight{} and \level{} was significant for all metrics (FC: \anovazh{1.0}{76.0}{4.636}{ = 0.034}{0.057}), FFD: \anovazh{1.0}{76.0}{6.977}{ = 0.010}{0.084}, TT: \anovazh{1.0}{76.0}{3.987}{= 0.049}{0.050}). This means the changes from \zero{} to \one{} are significantly different across \highlight{} method. 
The increase of fixation from \zero{} to \one{} for \rake{} is much higher than \gptsm{} (FC: 57.0\% higher, FFD: 43.6\% higher, TT: 60.4\% higher) as shown in figure~\ref{fig:highlight_guidance}, indicating that \rake{} highlighting guides reading better than \gptsm{}.


    
\paragraph{\textbf{Self-spoken vs. External text.}}

To understand the differences between reading self-spoken text and external text, we combined the five \interface{} conditions into one: \combined{} (all tokens from five \interface{} conditions), and compared it with \easy{} and \hard{}.
We performed Kruskal-Wallis tests for fixation metrics (FC, FFD, and TT) and found significant differences across \combined{}, \easy{}, and \hard{} for all three measures, as detailed in Appendix C, Table~\ref{tab:interface_external}. Pair-wise comparisons were performed with post-hoc Dunn's tests, which found self-spoken text \combined{} scored slightly lower than external text \easy{} and much lower than \hard{}, for all three fixation metrics (FC, FFD, and TT). 
A Kruskal-Wallis test was also performed for \regressionin{} across \combined{}, \easy{} and \hard{}. It only found a significant difference between \combined{} and \easy{} (p=0.009). 
\combined{} had the highest average \regressionin{} \meanse{13.73}{0.76}, followed by \hard{} \meanse{10.35}{1.33}), and then \easy{} \meanse{7.9}{0.84}.

These results clearly show the difference in reading effort between self-spoken (familiar) text and external (unfamiliar) text on fixation metrics and regression. As we can see in Figure~\ref{fig:regression_within} and \ref{fig:all_token_review}, self-spoken text had the highest regression and lowest fixation measures. One interesting side observation here: although we prompted the LLM to generate complex sentences for the \hard{} tasks, the generated text did not differ much in sentence structural complexity, but mostly adopted unfamiliar words to increase difficulty.

\paragraph{\textbf{Answers to RQ3.}}
In terms of \emph{reading effort}, we found that \as{} has a significantly lower inline regression rate than most other interfaces (\plain{}, \aoc{} and \rake{}), yet has significantly higher fixation measures (FC, FFD, TT) than them. While regression indicates \emph{difficulty} on the comprehension level \cite{sharmin2016reading}, fixation count and duration are more associated with word identification and \emph{unfamiliarity} \cite{rayner1998eye}. This means, although \as{} introduced unfamiliar words, it actually reduced the difficulty of comprehension. We believe \emph{unfamiliarity} was introduced by word substitution and paraphrasing when LLMs polished the original composition, which reduced \emph{difficulty} for reading by improving coherence and sentence structure. This disparity is interesting, because it is common for difficulty in word identification to accompany difficulty in comprehension \cite{rayner2006eye}. We can see this in our \easy{} and \hard{} external text baselines: \hard{} has higher regression and fixation than \easy{} (Figure~\ref{fig:regression_within}). This result suggests that an LLM-based abstractive gist extraction performs better than extractive highlighting methods for aiding the review of spoken text.

In addition, we were surprised to find that the \rake{} highlights could better guide reading than \gptsm{}. Our potential explanation is that \gptsm{} is designed for skimming \emph{unfamiliar written} text, its grammar-preserving function may not always work well for oral expressions with occasional grammatical mistakes. On the other hand, \rake{} might have served as effective signposts for self-spoken, familiar text. 

\begin{figure*}
\centering

\begin{subfigure}[b]{0.24\textwidth}
   \centering
   \includegraphics[width=\textwidth]{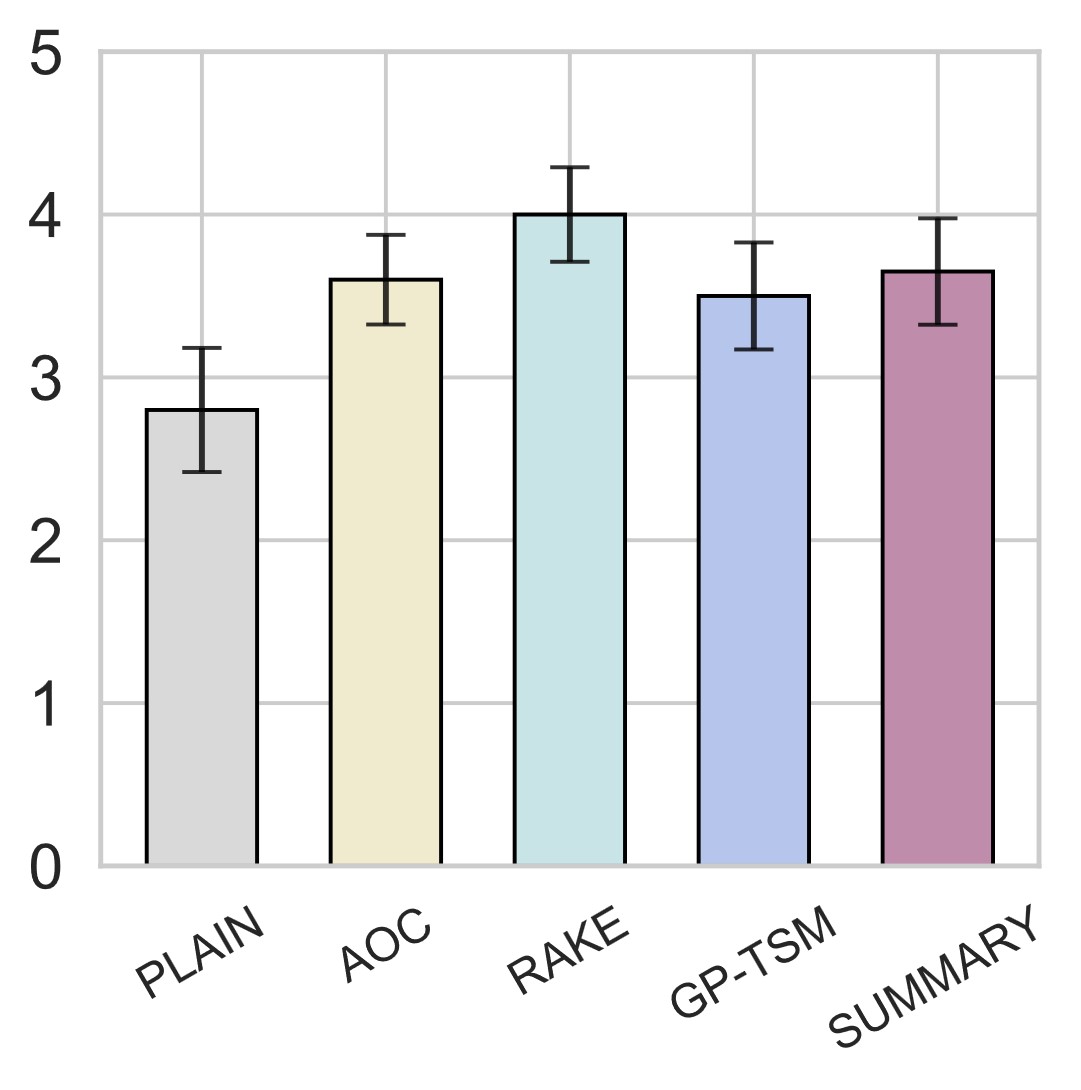}
   \caption{Preference in Speaking}
   \label{fig:rank_speaking}
\end{subfigure}%
\begin{subfigure}[b]{0.24\textwidth}
   \centering
   \includegraphics[width=\textwidth]{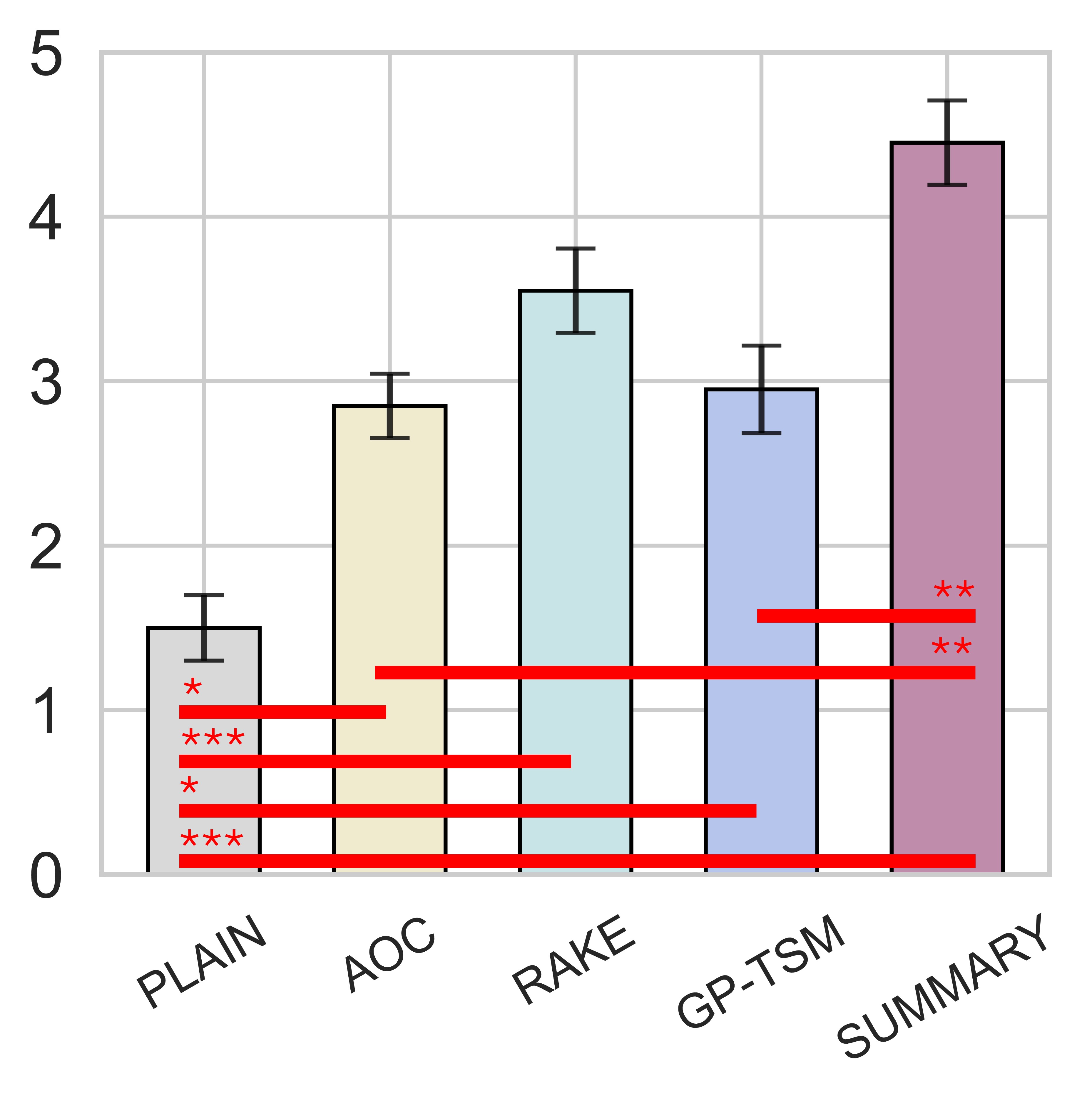}
   \caption{Preference in Reviewing}
   \label{fig:rank_reviewing}
\end{subfigure}%
\begin{subfigure}[b]{0.24\textwidth}
   \centering
   \includegraphics[width=\textwidth]{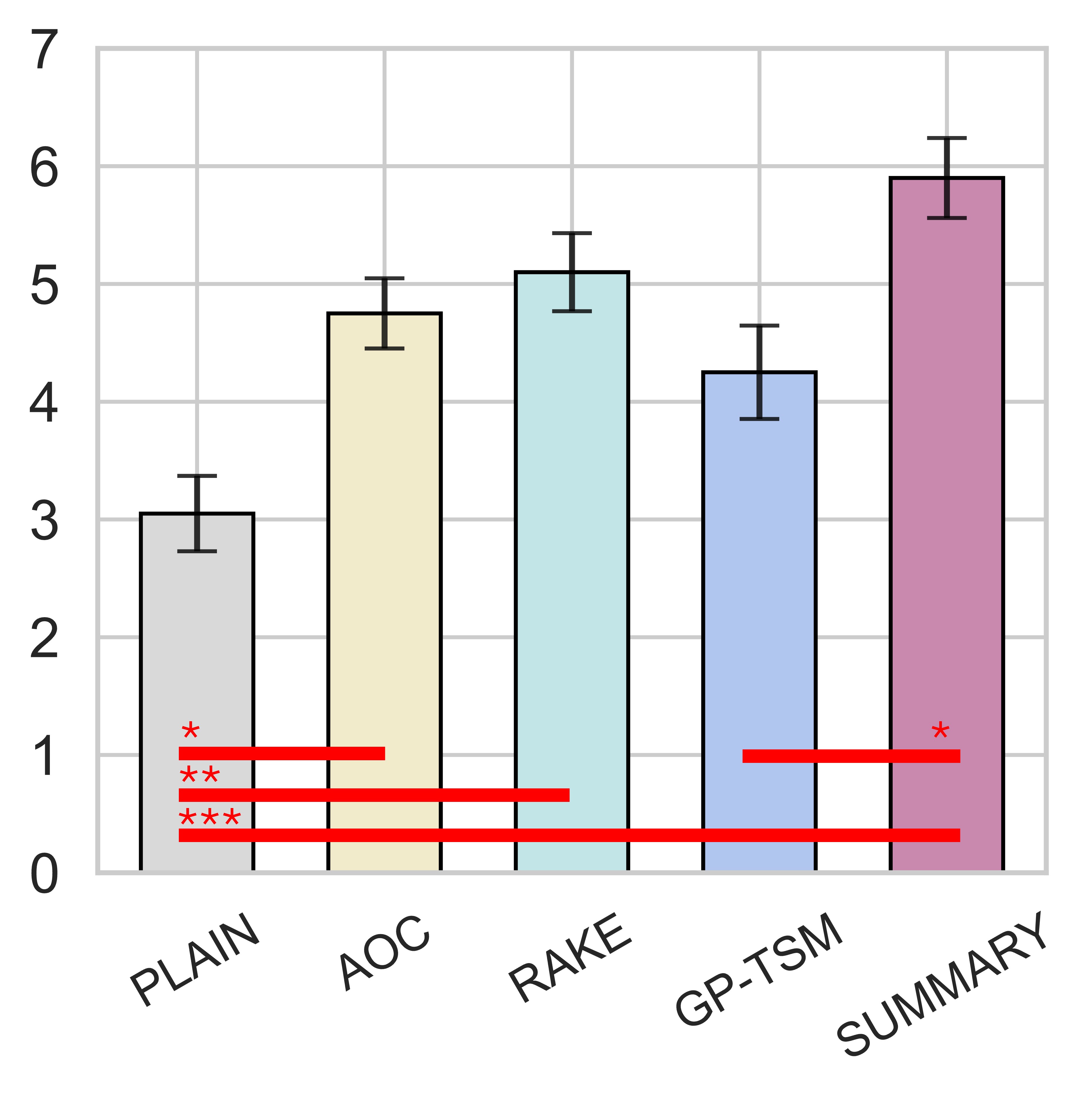}
   \caption{Content Satisfaction} 
   \label{fig:ques_satisfied}
\end{subfigure}%
\begin{subfigure}[b]{0.24\textwidth}
   \centering
   \includegraphics[width=\textwidth]{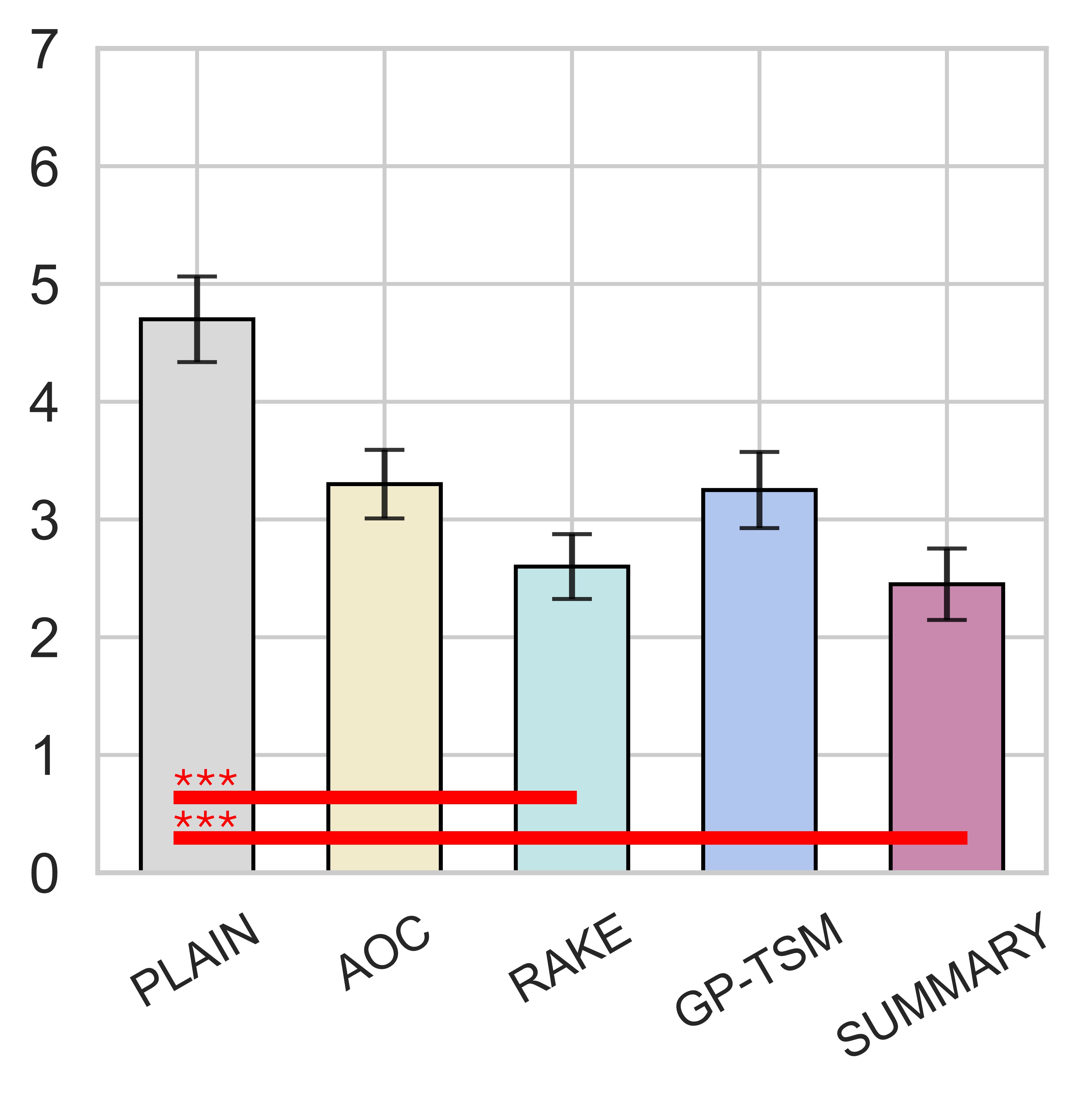}
   \caption{Mental Demand in Reading} 
   \label{fig:ques_frustration}
\end{subfigure}

\caption{Subjective rank-based scores of helpfulness in speaking (a) and reviewing (b), and questionnaire 1-7 likert scale question scores for “I am satisfied with the content in this transcription” (c) and “How mentally demanding was it to read the transcript” (d)} 

\label{fig:questionnare_and_ranking}
\end{figure*}

\section{Subjective feedback}
Participants were asked to rank their preferences during the speaking and reviewing phases respectively after completing all the tasks. These rankings were converted into scores for quantitative analysis, with the first rank receiving 5 points, the second 4 points ... the last 1 point. They were allowed to give equal ratings for different conditions. In addition, we asked them to rate their content satisfaction and frustration in reading. 
In this section, we summarized our findings from surveys and semi-structured interviews and aimed to answer the fourth research question. \\




\subsection{RQ4: Which interface(s) do participants prefer?}

\textbf{\rr{User Preference.}} Some participants found it hard to choose their preferred interface for the speaking process. The most preferred top choice was \rake{}, with six participants favoring it, and the rest of the top votes spread across other conditions. 
For the preference in reviewing, the most preferred top choice was \as{}, with 15 participants favoring it, followed by \rake{} with four participants favoring it. 
After converting all the rankings into scores, Figure~\ref{fig:questionnare_and_ranking} showed the results for the speaking and reviewing phases, respectively. 
For the speaking phase, a Kruskal-Wallis test found no significant difference across \interface{} (p=0.200). This is consistent with the quantitative results that also found no significant difference across interfaces during speaking, probably because they spent little time reading.

For the reviewing phase, a Kruskal-Wallis test identified a significant difference across \interface{} \kruskalsse{4}{46.14}{ }{.444}. As shown in Figure~\ref{fig:questionnare_and_ranking}-(b), \as{} received the highest average score, with significant differences from \plain{} (p<0.001), \aoc{} (p=0.002), and \gptsm{} (p=0.006), but not \rake{} (p=0.381). \plain{} received the lowest score, which was also significantly different from \aoc{} (p=0.043), \rake{} (p<0.001),  \gptsm{} (p=0.017), and \as{} (p<0.001). We found no significant difference across \aoc{}, \rake{} and \gptsm{} themselves. In summary, during the reviewing process, participants mostly preferred \as{}, then \rake{}. Then they ranked \aoc{} and \gptsm{} rather similarly. \plain{} was the least preferred condition for reviewing.



\textbf{\rr{Content Satisfaction.}} Figure~\ref{fig:questionnare_and_ranking}-(c) showed the ratings of how satisfied they were with the content being created for each interface. 
A Kruskal-Wallis test identified a significant effect for content satisfaction across \interface{} (\kruskalsse{4}{30.31}{}{.277}). Dunn’s post-hoc test showed that \plain{} was rated significantly lower than \aoc{} (p=0.037), \rake{} (p=0.003) and \as{} (p<0.001). Moreover, \as{} was rated significantly higher than \gptsm{} (p=0.014). Although the content of \rake{} and \gptsm{} is technically just \aoc{} adding highlights, the highlights seemed to have affected the rating for content satisfaction as well. \gptsm{} was ranked as the second lowest, above \plain{} and below \aoc{}, showing that the grammar-preserving highlighting did not work well for spoken content. 

\textbf{\rr{Mental Demand in Reading.}} Figure~\ref{fig:questionnare_and_ranking}-(d) showed the results of the rating in the Kruskal-Wallis test were significant \kruskalsse{4}{22.97}{}{.197}, indicating a large effect across conditions. Dunn’s post-hoc test showed that \plain{} was rated significantly higher than \rake{} (p<0.001) and \as{} (p<0.001). 
We did not find a significant difference across the other four interfaces besides \plain{}. This indicates that all four interface interventions tested in our experiment had a similar perceived mental load, which is lower than the raw transcript.

\subsection{User Experience}
In the interview with participants, we asked for subjective feedback for each interface condition. We synthesize their experience from the dimensions where our interface conditions differ: the fidelity to the original speech, the effectiveness of methods for aiding review, and the interface's responsiveness and stability during speech production.

\subsubsection{Fidelity to Original Speech}
Participants' feedback revealed a clear trade-off between the fidelity of the transcription and its final quality.

Regarding the \plain{} interface, participants found that speech recognition and punctuation errors were distracting and negatively affected reading. Seven participants were displeased with excessive punctuation. P8 explained, \inlinequote{I find it frustrating when the system places periods in strange places. I thought I was taking a break to think, and my sentence hadn't finished, but the system put a period and started a new sentence, which made me very tense, so I tried to speak faster.} During the review phase, eight participants were dissatisfied with the transcription quality. P1 remarked that \inlinequote{it has a lot of errors or grammar mistakes,} and P19 noted that \inlinequote{some specific words and keywords are not well recognized, affecting the understanding of the text.}

In contrast, the \aoc{} interface was praised for striking a balance that improved fluency while preserving the original meaning. Five participants expressed that they appreciated its corrections. P3 noted, \inlinequote{The sentences are more complete, and the spoken text has become readable text. Compared with \plain{}, it is not so fragmented.} P9 mentioned that the system \inlinequote{Deleted some useless words and a lot of my repetitive language habits.} Crucially, participants felt their intent was maintained, as P14 stated, \inlinequote{The original meaning was not changed, and my mistake was corrected. My speech was expressed truthfully.}

The \as{} interface was perceived as concise and well-organized, although its heavy processing risked altering the original meaning. Thirteen participants praised its ability to turn spoken language into formal written text. P8 remarked, \inlinequote{It's very clever and good at organizing things. The reading process is the simplest... The word choice is very accurate.} P19 remarked, \inlinequote{It no longer sounds like speech but more like written language, even feeling like writing an article.} However, this also came at the cost of potential misinterpretation. P11 noted, \inlinequote{Although it writes more precisely, sometimes it conveys a different meaning from mine. It misinterprets.} 

\subsubsection{Methods for Aiding Review}
The interfaces employed two distinct strategies to aid text review: highlighting important content and improving overall readability.

For highlighting, the keyword-based strategy in \rake{} was considered effective for speeding up reading, though some found highlighted keywords scattered. Nine participants expressed that \rake{} could improve reading speed. P4 believed that \inlinequote{Highlighting helped me actually go through the text a lot faster,} and P5 mentioned, \inlinequote{It reduces the reading time as the eyes scan the highlighted words.} However, the scattered nature could negatively affect reading, as P3 remarked, \inlinequote{It's very scattered; words are everywhere, making it hard to piece together. I think it does not help my reading.}

In contrast, participants found the phrase-based highlighting of \gptsm{} to be distracting and often too excessive to be effective. While four participants said it could help capture main information, nine reported negative experiences. P8 noted, \inlinequote{There's too much emphasis, making the information hard to process.} 

Moreover, improving the text's fundamental readability was a highly effective method for aiding review. The light corrections of \aoc{} were a clear improvement over \plain{}, as P3 noted the text was \inlinequote{not so fragmented.} The summaries from \as{} were particularly praised for their clarity. P4 commented, \inlinequote{The summary is really easy to read because the sentences are well-organized.} P8 added, \inlinequote{After the summary, it is very easy and comfortable to read.}

\subsubsection{Interface Responsiveness and Stability}
Participants' real-time experience during speech production was heavily influenced by the interface's responsiveness and stability.

Participants appreciated real-time feedback but found processing delays in complex interfaces could create anxiety. The \plain{} interface was valued for its speed, as P9 stated, \inlinequote{It is the most intuitive, and delays in other tasks disturb me, affecting my input.} 

Similarly, drastic or constant visual changes on the interface were seen as distracting and could disrupt the speaking flow. Three participants commented that \gptsm{} was overly distracting. P4 commented, \inlinequote{It made the screen a lot more chaotic and full.} The changes in \as{} were also described as disruptive. P4 mentioned, \inlinequote{The changing stopped me from talking,} and P12 noted on the visual distraction caused by the text as \inlinequote{It suddenly gets shortened or lengthened.}

Conversely, subtle and helpful feedback could enhance the speaking experience by boosting confidence or providing a sense of safety. Two participants (P7, P17) elaborated on how highlights in \rake{} could boost their confidence. P17 noted, \inlinequote{When I say a sentence with a key point, I want to see if this point has been recorded. If it captures and highlights this point, it helps me to see and confirm this matter.} For \as{}, the organizing capability provided a sense of security. P16 expressed, \inlinequote{I feel safe. No matter what I say, it will be able to organize it... I don't have too much fear of losing memory when I speak, because the summary will help me round it up.}

\subsection{Explanations of Gaze Engagement During Speech Production} 
\label{sec:gaze_explained}

In the interview, we asked participants to reflect on their gaze behaviors during speech production and talk about reasons for looking or not looking at the text. We summarized our findings below. 

\subsubsection{Avoiding distractions.} Nine participants 
reported they intentionally avoided looking at the interface to avoid distraction. P1 remarked, \inlinequote{(During speaking,) I look (at the text), but I don't read because I'm occupied by speaking. If I started reading, then I get confused or distracted.} P10 stated, \inlinequote{I don't pay much attention to my vision, but more to what's going on in my mind.} P13 explained, \inlinequote{If I look at the interface, I would want to align my speaking speed with the transcription. But this would disrupt my thinking rhythm.} This helps explain the low percentage of \sustained{} found in Section  \ref{chap:eq1_users_visually_engage}.

\subsubsection{Reasons for active looking}
Six participants 
mentioned they were mainly monitoring the system operation or task progress. P6 stated, \inlinequote{I only looked at the interface to see how much I was speaking and how much I still have left to speak. So like to check on the progress.} P11 explained, \inlinequote{The reason I would look at it was when there was a sudden change, such as more text. 
Or after speaking several words, it still didn't show. I would check the operating status of the system.} P7 mentioned, \inlinequote{I check whether the content is accurately transcribed. Sometimes I will wait for the text to come out before I start to say the next sentence.}

Four participants 
reported unconsciously glancing at the interface without a specific purpose. P15 stated, \inlinequote{I would look at the interface, mostly unconsciously, without any purpose.} P12 mentioned, \inlinequote{My eyes may glance around, but I don't read it.} This helps explain part of the \hopping{} gaze patterns we found in Section \ref{chap:eq1_users_visually_engage}.

\subsubsection{Reasons for passive looking}
Six participants 
noted that changes in the interface attracted their attention. P4 remarked, \inlinequote{When I was speaking and the AI was changing the sentence structure, that made me look into the interface.} P8 stated, \inlinequote{Some words that pop up by highlights would draw me in. Not a change in meaning, just a visual change.} 


Three participants
mentioned being drawn to transcription errors. P12 noted, \inlinequote{Punctuation marks will also attract me if they are very different from what I think.} 
P18 mentioned, \inlinequote{I said something wrong, and the mistakes were also transcribed.}

\section{Discussion}

\toc{Our eye-tracking study of dictation interfaces uncovered novel insights into users' visual engagement with spoken text and their reading effort during speech production and review. Below we summarize our findings, discuss them in the context of existing knowledge, and explicate the implications for design. }

\subsection{Findings and Future Research Directions}

\subsubsection{Sparse Reading in Speech Production} 
We found that gaze engagement patterns differ substantially during speech production and reviewing processes. \toc{Our participants spent only 7-11\% of the time reading text during production. This has an interesting consistency with a previous study on written composition, where participants spent only 5.8\% of their time in sustained reading \cite{torrance2016reading}. Language production likely occupies the majority of the cognitive load during production, regardless of the input modality, leaving little cognitive resources for sustained reading. In our study, the participants' gaze engagement during composition was explained in their subjective feedback: they looked away to avoid distraction while speaking; they were mainly monitoring the system and task progress while intentionally looking; their eyes might also be attracted by interfaces and errors. We found these gaze strategies are highly persistent, not affected by the interfaces displaying / altering the produced text. This could inform future gaze-aware systems for detecting users' status and providing adaptive UI support. Future work could explore effective interface design for progress monitoring while reducing distractions when the user is speaking. Moreover, a potential research direction is to explore graphical representations of the gist of the spoken text. For instance, it could provide an overview of the topics or visualize the evolvement of the semantic space. Figure~\ref{fig:design_implication}-a and b illustrate example design concepts for these purposes. }  



\subsubsection{Desirable Unfamiliarity}
\toc{One of the most striking findings is the phenomenon we term ``\emph{Desirable Unfamiliarity}'': abstractive summaries, despite introducing unfamiliar words, reduce reading difficulty and are preferred by users. This suggests that while word identification may become slightly more challenging due to unfamiliar phrasing, the overall coherence and readability of the text improve, leading to fewer regressions and easier comprehension. This finding challenges the conventional wisdom that familiarity with text always leads to easier reading \cite{leroy2014effect, rayner1998eye}. 
Our finding also highlights the importance of prioritizing gist comprehension over verbatim accuracy in dictation interfaces, in line with previous speculation in the literature \cite{10.1145/3571884.3597134}. Participants' most favored interface being the LLM summaries indicates a level of acceptance towards altered wording from their original speech as long as the main message remains intact. This opens up new research questions on understanding the effects of different summaries: how much deviation it could depart from the original transcript to be still acceptable, which could include both their conciseness and the generation strategies customized for various types of content. }




\rr{Besides the potential memory effect of self-spoken content, another possible explanation for the desirable observation of unfamiliarity comes from the nature in which LLMs compose the text repeatedly predicting the next word \cite{brown2020language}, forming a predictable and coherent sentence even though the individual words could be unfamiliar to the reader. This is inline with the results of a recent work \cite{ilyas2025reading} which suggests that ``AI generated texts could be more convenient to read" and "current LLM models are generating text with ... simplified structures." Of course, more studies are needed to replicate and further explore the desirable unfamiliarity hypothesis.}


\subsubsection{Highlighting for Recall}
Finally, the abstractive information extraction method (LLM summaries) outperformed extractive summaries (highlights) in reading one's spoken content. Grammar-preserving extractive summaries were known to be effective for assisting reading unfamiliar text \cite{gu2024ai, gu2024skimmers}. We found it did not perform better than simple keyword spotting (RAKE) in our context. \toc{This indicates that text highlighting strategies for self-produced (familiar) text shall focus on sign-posting for facilitating recall rather than comprehension. Future research could explore effective sign-posting methods through keyword highlighting, e.g., special nouns, important verbs or conjunction words. Personalization methods could also be explored, as the individual differences in preferred memory support could be large. } 

\begin{figure*}
    \centering
    \includegraphics[width=\linewidth]{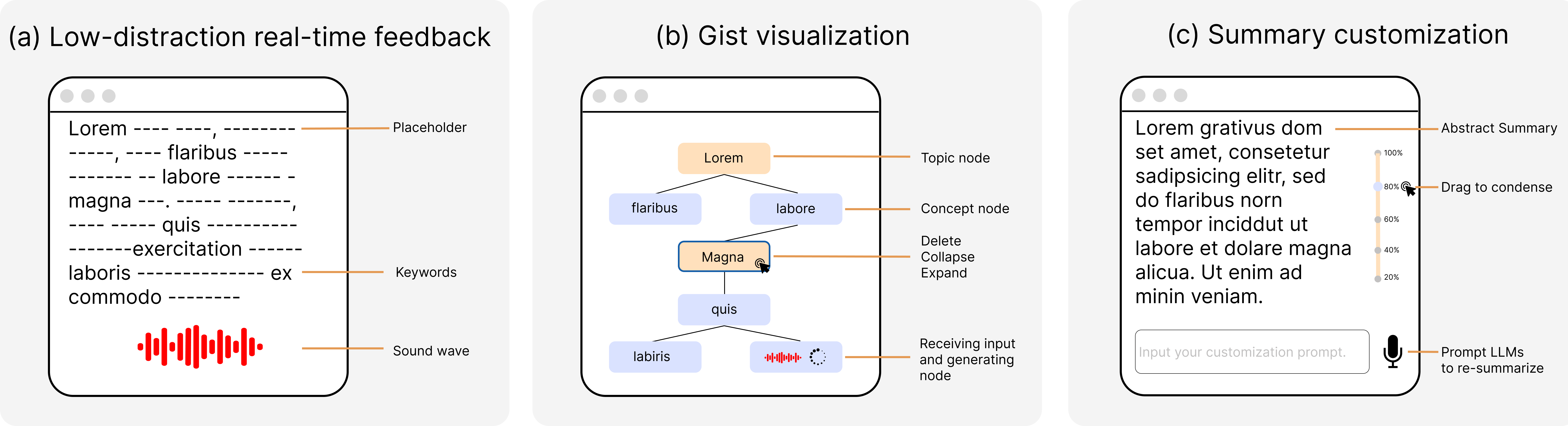}
    \caption{Based on our findings, we illustrate three examples of interface concepts for future dictation tools: (a) Low-distraction real-time feedback, displaying only keywords and placeholders to reduce distraction while keeping real-time feedback (also with soundwave animation) about the speech input and recognition progress. (b) Gist visualization, moving beyond only textual feedback about the spoken content, by structuring the text into a concept map to facilitate review and manipulation. (c) Summary customization, generating summaries of varying conciseness with an option to customize them via user prompts. These designs aim to reduce cognitive load while providing necessary visual support, help users clarify core ideas (gist), and offer ways to tune summaries.}
    \label{fig:design_implication}
\end{figure*}

\subsection{Design Implications}


\subsubsection{Real-Time Feedback \& Distraction Reduction} During speech production, users spend little time reading the spoken text. This finding redirects the attention of interface designers away from improving the readability of text to reducing unnecessary distractions in production. \toc{Perhaps there could be different interface solutions supporting production and review \cite{10.1145/3542922}. For example, interfaces could provide subtle visual cues to indicate transcription progress without overwhelming the user with text changes, a conceptual interface is shown in Figure~\ref{fig:design_implication}-(a). Previous work on eyes-free dictation and editing \cite{ghosh2018editalk} showed a very high cognitive load due to missing visual support. Therefore we believe the balance of visual demand is a key design challenge. Existing adaptive user interfaces for reducing distractions based on users' cognitive load, eg. in driving context \cite{wen2024adaptivevoice, tchankue2011impact}, could be one direction to explore. }

\toc{\subsubsection{Extract Gists Instead of Correct Errors}
The development of STT technology has been much focused on correcting recognition errors \cite{ERRATTAHI201832}. Our study showed that only correcting speech recognition errors is not enough to support dictation.  
We were surprised that \aoc{} did not perform significantly better than \plain{} for either regression or fixation metrics, as we confirmed from participants' subjective feedback that \aoc{} effectively improved the recognition accuracy. While future research is needed to dive into this, we suspect two potential causes. One is related to the participants' speech production ability. Although \aoc{} helps correct recognition errors, it does not fully correct the human errors made in speaking, such as wrong wordings, which would also affect reading. Second is the nature of oral language -- the choices of wording and sentence structures might be different from written language, eg. being more verbose, which could also negatively affect the reading. Therefore, we advocate for a more holistic approach to improving transcripts beyond correcting disfluencies and punctuation. }


As discussed, this work showed the effectiveness of displaying summaries to support the review of spoken text. However, higher-level summaries necessarily compromise details and become harder to align with users' intentions. One solution of ``tuning'' summaries was introduced in Rambler \cite{lin2024rambler} by toggling keywords and regenerating, a conceptual interface is shown in Figure~\ref{fig:design_implication}-(c). The effectiveness of this and its future improvements need to be studied further. Furthermore, going beyond textual summaries, many graphical methods could be used in gist representation, such as mind maps, concept maps, storyboards, infographs as well as other more sophisticated visualizations~\cite{el2017interactive, gold2016visargue}, a conceptual interface is shown in Figure~\ref{fig:design_implication}-(b). 

\subsubsection{Go Beyond a Text Editor Interface} Existing dictation interfaces, such as the ones provided by Android or iOS systems, inherit most foundations from the traditional text editor interfaces. As a typewriting interface, its interaction happens almost solely around the cursor area. 
As our participants spend only half of the gaze-on-text time looking at the immediate production, we suggest that the design of dictation interfaces should not only focus on the cursor area, but also the entire content display. Future research could consider investigating other potential supports during composition instead of reading. Recent LLM-supported writing interfaces are already moving the writing task away from word-to-word composition to piece-by-piece manipulation \cite{buschek2024collage}. When it comes to dictation, we could consider providing visual cues for monitoring progress and helping plan the composition\cite{clark2018creative, osone2021buncho, 10.1145/3290605.3300526}. Effective highlights could help them recall the gists of previous production \cite{lin2024rambler}. In addition, users could be looking at other materials / objects while speaking when composing relevant content, eg. commentaries or reviews~\cite{yang2025ramblerwilddiarystudy, 10.1145/3453988}. Such tasks can also be better supported. 

\toc{
\subsubsection{Highlighting Strategies} While both RAKE and GP-TSM highlighting methods were effective, RAKE was more successful in guiding users' attention during review. This suggests that keyword-based highlighting may be more suitable for self-produced text. 
Highlighting as signposting for recall appears to be more important than reading comprehension. This leads to potential future development of customized highlighting strategies for STT interfaces. Perhaps there could be a better approach combining summaries and highlighting. However, interface designers need to be mindful of the distraction it may cause in the speaking process if the text visually changes too drastically. Perhaps smoothing animations can be developed to mitigate the sudden jumps caused by word and length changes. Going beyond keyword highlighting, other visual highlighting methods could be explored, including color combinations and motion \cite{10.1016/S0378-7206(02)00091-5}. The interventions for assisting low-vision reading can be considered as aids here as well \cite{Sandnes2025}. More graphical solutions for extracting gists of content can be explored, such as automatic concept map generation \cite{10.1145/3544548.3581260, lee2012knowledge}. 
}




\toc{
\subsubsection{Broader Implications}
Our findings extend beyond dictation interfaces and have implications for other applications where transcribed content is displayed, such as AI-summarized meeting notes, lecture transcripts, and conversational interfaces. Recent works in other research areas have explored ways to handle STT transcripts such as creating structured medical notes from audio recordings\cite{kernberg2024using}, which highlighted the inaccuracy in summarization as a major limitation in healthcare. Hakulinen et al. \cite{hakulinen2021design} formalized how different professional areas require different styles of voice-based reporting, including how real-time it needs to be and how transcriptions are summarized and structured. 
Our findings, such as the concept of \emph{Desirable Unfamiliarity}, could also inform the design of these systems, 
given these applications also work with familiar text although not always self-produced. 
}

\subsection{Limitations}
Our work has several limitations. First, \cam{due to the lack of control over LLM output, the text update in \gptsm{} and \as{} conditions can be unstable sometimes. Although our diverse sampling strategy partially mitigated this issue, it did not completely address it.} The delay caused by the API calls for offline correction and LLM processes could also affect user experience. We partially controlled for this issue by keeping the delay the same across four conditions. This will become less of a problem in the future as AI models continue to upgrade. 

\cam{A low-frequency eye tracker like ours (60hz) has limited temporal accuracy. However, this usually affects the reliable study of short eye movements such as saccades or micro-fixations. Our case is less affected as we focused on fixations longer than 50 ms, which is typical in reading research to capture cognitive processing. The eye-tracking correction algorithm we used further increased data reliability and reduced this limitation (Figure~\ref{fig:drift_correction}).}

\cam{Our study used a social media composition task. As noted in writing research, such public-facing tasks create an ``envisioned audience'' \cite{litt2016imagined} compared to personal writing. While it served our purpose of enforcing a careful read of the passage, this context could have an influence on users' acceptance of text alterations by LLMs. Future research should explore such effects. Additionally, different from other reading studies using text given by the experimenter, our text composition task makes the reading material different in every trial for every participant, which creates an unavoidable source of noise in the data. We believe this is partially why our statistical findings for the fixation metrics have rather small effect sizes. }

\cam{We acknowledge the limitation of a study sample consisting primarily of university students. Future studies should recruit participants with more diverse backgrounds.} While the state-of-the-art STT tools primarily transcribe English, we conducted the study in a non-English-speaking region, which limited our ability to recruit native speakers of the tested language. Language fluency shall play a role in the quality of production, which subsequently affects reading. Our analysis of native/bilingual English speakers' experience is only preliminary due to the limited sample size. \cam{However, it introduces questions and suggests that native speakers may react differently to the interfaces (e.g., issues in raw transcripts) from non-native speakers.} Future research could replicate the study with native speakers to better understand this.
\section{Conclusion}
To our knowledge, this is the first eye-tracking study investigating gaze behaviors in speech production and evaluating the reading experience of self-spoken text. Our findings uncovered insights into visual engagement patterns and reading effort in terms of difficulty and unfamiliarity. We conclude that LLM-generated summaries can effectively reduce the reading cost of spoken text despite introducing unfamiliarity. 
Dictation interfaces should prioritize supporting recall and might benefit from presenting gist-level representations—such as summaries or other abstractions—rather than maintaining strict verbatim fidelity as in traditional text editors. At the same time, system designers must remain attentive to preserving the original intent of the composition, potentially through new design strategies and technical solutions.
These findings can inform pivotal design decisions of STT interfaces, which are widely accessible as alternative text input and recording methods today.  



\begin{acks}
This research is supported by a donation (project No. 9220140) from Huawei Technologies Co., Ltd., and the Google Faculty Research Award (project No. 9229068) from Google LLC.

We acknowledge the use of ChatGPT in refining wordings and making text more concise for Section 7.2 and Section 8, with careful review by the researchers to make sure the original meanings remain unchanged. The authors take full responsibility for the correctness and authenticity of the content. 
\end{acks}

\bibliographystyle{ACM-Reference-Format}
\bibliography{main}

\clearpage 
\onecolumn 
\appendix

\section{Native/Bilingual English speaker participants' fixation metrics in reviewing process.}
\label{appendix_native}
\begin{figure}[h]
\centering

\begin{subfigure}[b]{0.33\columnwidth}
   \centering
   \includegraphics[width=\textwidth]{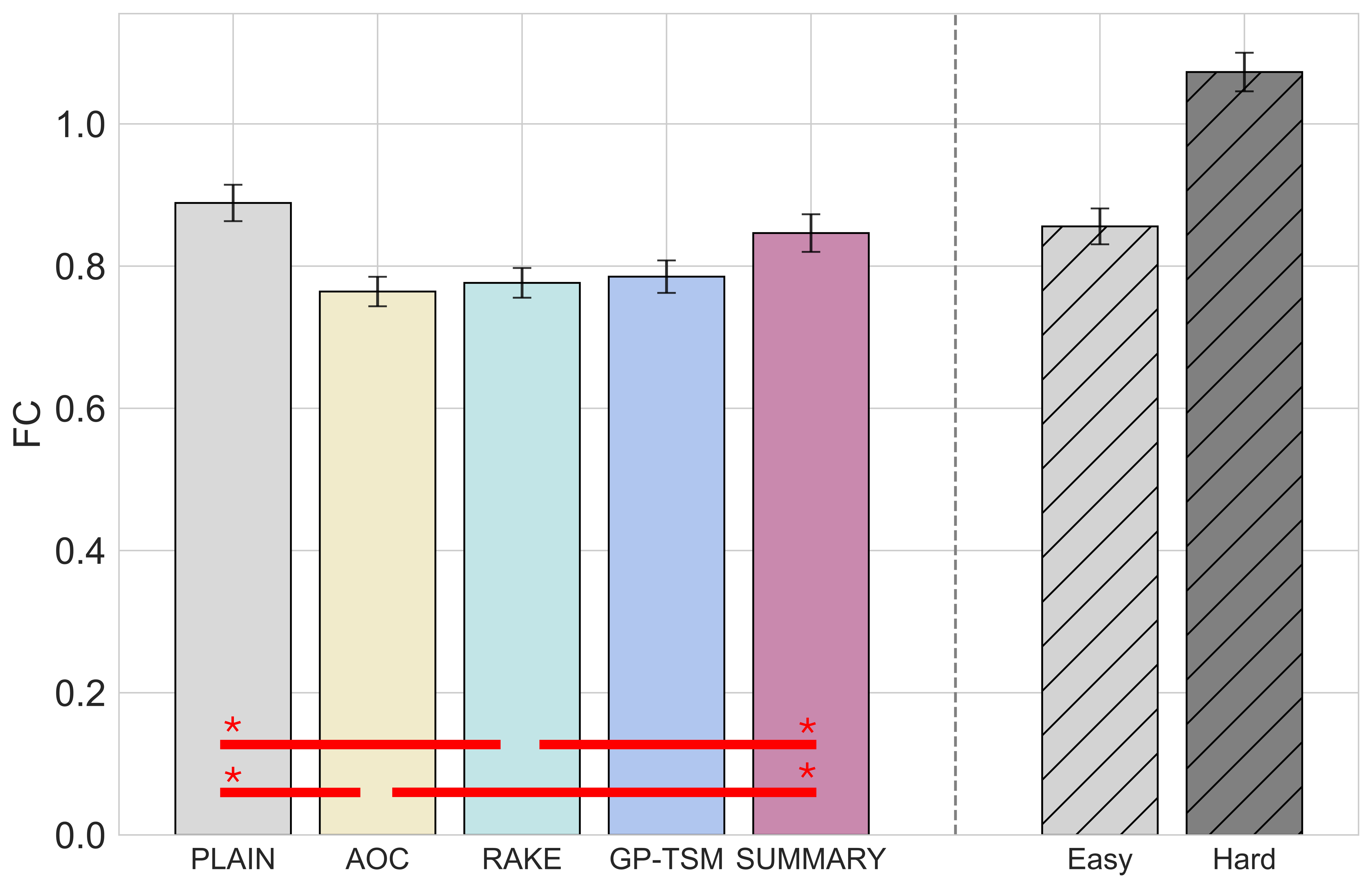}
   \caption{Fixation Count (FC).}
   \label{fig:all_fc}
\end{subfigure}
\begin{subfigure}[b]{0.33\columnwidth}
   \centering
   \includegraphics[width=\textwidth]{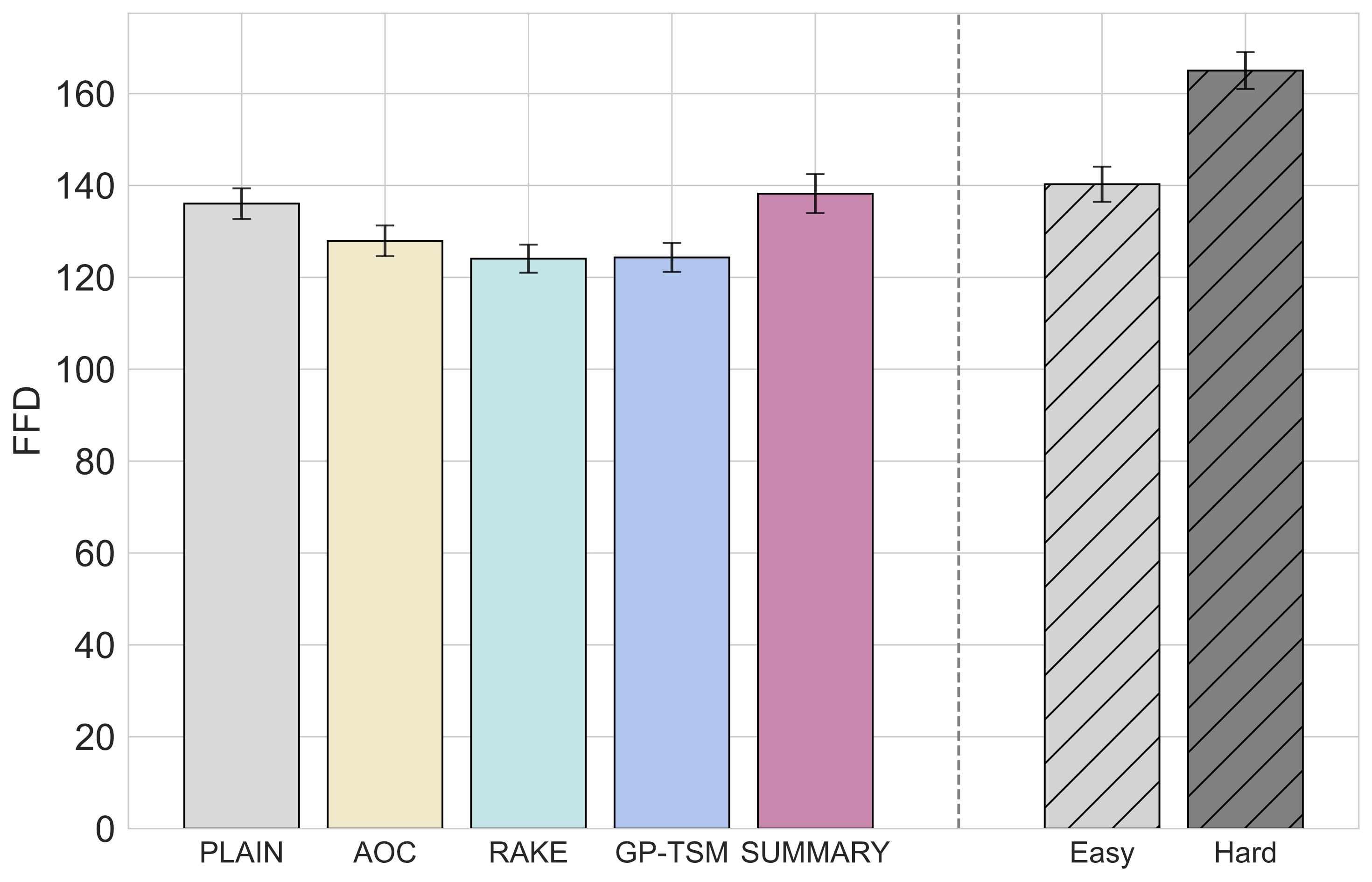}
   \caption{First-Fixation Duration (FFD).}
   \label{fig:all_ffd}
\end{subfigure}
\begin{subfigure}[b]{0.33\columnwidth}
   \centering
   \includegraphics[width=\textwidth]{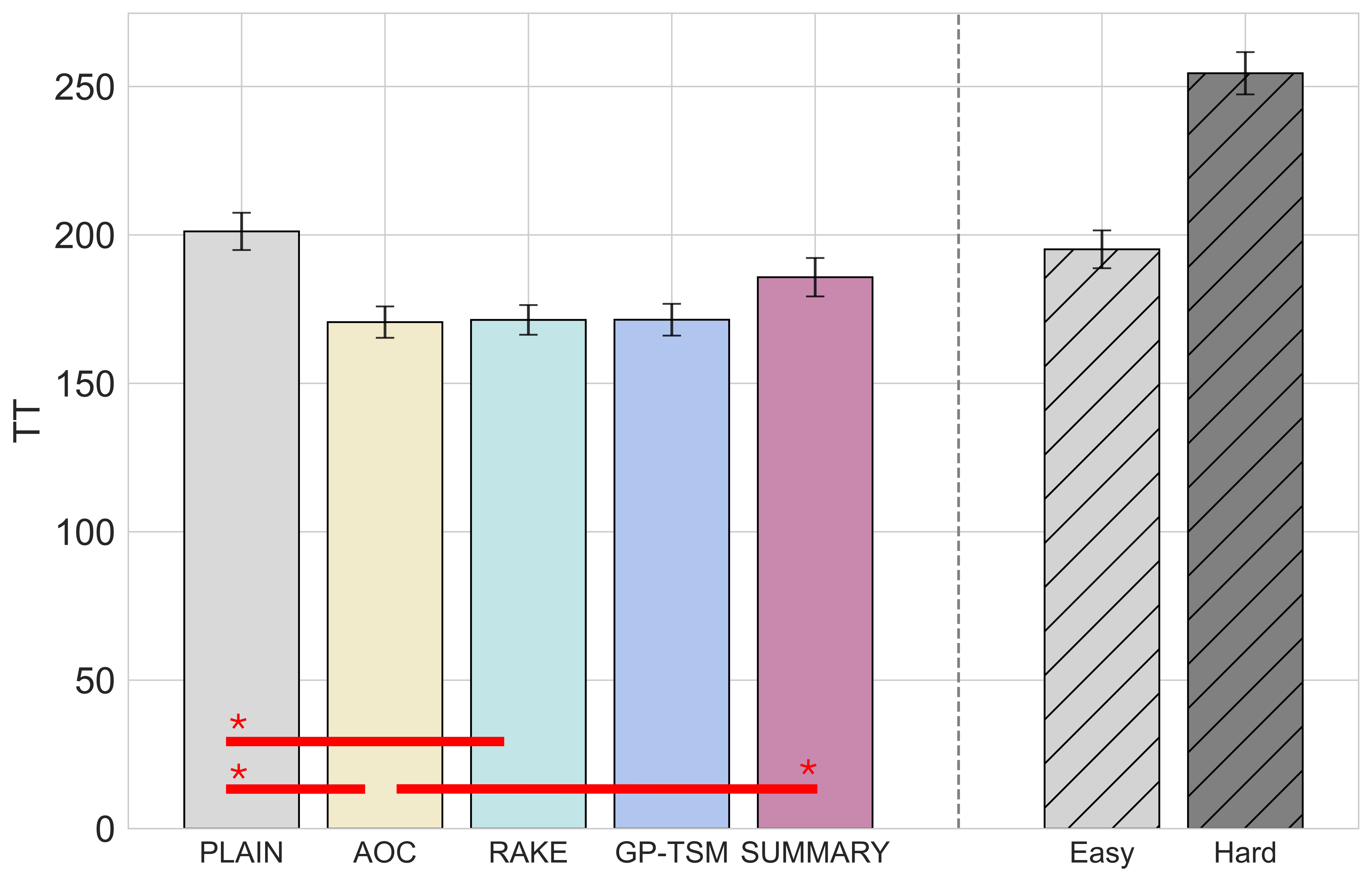}
   \caption{Total Time (TT).}
   \label{fig:all_tt}
\end{subfigure}

\caption{Native/Bilingual English speaker participants' fixation metrics in reviewing process (duration in milliseconds). We analyzed the same token-level metrics (fixation) on a subset of the data only from the six native/bilingual English speakers.
Kruskal-Wallis test show significant differences across five \interface{} conditions for all three metrics listed here. Specifically, 
    FC: \kruskalsse{4}{23.01}{}{.002}, 
    FFD: \kruskalnse{4}{13.38}{.010}{.001}, 
    TT \kruskalsse{4}{22.73}{}{.002}. 
Dunn’s post-hoc test showed that for FC and TT, \plain{} is significantly higher than \aoc{} and \rake{}. \as{} is also significantly higher than \aoc{} and \rake{} for FC, and significantly higher than \aoc{} for TT.
We did not find significant differences in FFD across \interface{} conditions. Only significant differences between \interface{} conditions are marked on the figure to avoid cluttering.} 
\label{fig:native_token_review}
\end{figure}

\section{\rr{Example Content in External Reviewing Tasks}}
\label{easy_and_hard_example}
\easy{} Example: Hey everyone! Just wanted to drop a quick note about our spontaneous road trip this weekend. We decided to shake things up and drove to the serene Lakeview Park. It was absolutely the breath of fresh air we needed! The journey started with a sunny drive, music blasting, and lots of laughter. Once we arrived, the view of the lake, with its crystal-clear waters, was stunning. We spent the day hiking around the area, enjoying the lush greenery and the peaceful environment. It was a great way to reconnect with nature and each other. We also had a picnic by the lake with sandwiches, chips, and homemade lemonade which made the day even more perfect. Watching the sunset over the lake was the highlight of the trip. It was one of those moments you wish you could freeze in time. I encourage all of you to take a little break from your daily routine and explore the beauty around you. It doesn't have to be far or expensive. Sometimes, the best spots are just a drive away! Can't wait for our next little adventure. Until then, keep exploring and stay happy!

\noindent \hard{} Example: I am excited to initiate a groundbreaking research project aimed at deciphering the complex mechanisms of synaptic plasticity in neurodegenerative diseases, with a focus on Alzheimer's. Utilizing advanced techniques like single-cell RNA sequencing and computational modeling, our goal is to uncover the interplay between genetic and environmental factors influencing synaptic dysfunction. Our interdisciplinary team will integrate molecular biology, neuroscience, and bioinformatics to map gene expression patterns and identify potential therapeutic targets. This research will not only deepen our understanding of Alzheimer's pathology but also propel personalized medicine by developing targeted treatments. I look forward to engaging with peers in this challenging yet promising domain, as we strive to innovate and uncover new therapeutic avenues against neurodegeneration.

\section{Tables from Statistical Asnalysis}
\label{appendix_table}
\begin{table}[htbp]
    \centering
    \caption{\rr{Analysis of task order effects. Each task was assigned a chronological order position (1–5) as the independent variable to evaluate its impact on the dependent variables (trial level metrics). The table summarizes the test statistics and $p$-values indicating the order of task did not significantly affect participant eye-movement metrics.}}
    \label{tab:ordering_effect}
    \begin{tabular}{lllc}
        \toprule
        
        \textbf{Stage} & \textbf{Metric} & \textbf{Test} & \textbf{\textit{p}-value} \\

        \midrule
        
        \multicolumn{4}{l}{\textit{Speech Production}} \\
        & \offtext & Kruskal-Wallis & 0.714 \\
        & \hopping & Kruskal-Wallis & 0.905 \\
        & \sustained & Kruskal-Wallis & 0.453 \\
        & \hopping~\distal & Kruskal-Wallis & 0.361 \\
        & \hopping~\local & Kruskal-Wallis & 0.613 \\
        & \sustained~\distal & Kruskal-Wallis & 0.913 \\
        & \sustained~\local & Kruskal-Wallis & 0.904 \\

        \midrule
        \multicolumn{4}{l}{\textit{Reviewing}} \\
        & \offtext & Kruskal-Wallis & 0.063 \\
        & \hopping~(\ontext) & ANOVA & 0.821 \\
        & \sustained~(\ontext) & Kruskal-Wallis & 0.509 \\
        
        & \regressionin & Kruskal-Wallis & 0.653 \\
        & \regressionbt & Kruskal-Wallis & 0.434 \\

        \bottomrule
    \end{tabular}
\end{table}
\begin{table*}[htbp]
\centering
\caption{\rr{Normality tests and main test selection for measures. ($p < 0.05$ indicates a violation of the assumption. ANOVA is used only when both Normality and Homogeneity are True; otherwise, Kruskal-Wallis is used.)}}
\label{tab:Normality_test}
\resizebox{\textwidth}{!}{
\begin{tabular}{llccccccc}
\hline
\textbf{Category} & \textbf{Measure} & \multicolumn{3}{c}{\textbf{Shapiro-Wilk (Normality)}} & \multicolumn{3}{c}{\textbf{Levene (Homogeneity)}} & \textbf{Main Test} \\ 
 &  & \textbf{$W$} & \textbf{$p$-value} & \textbf{Normality} & \textbf{Stat} & \textbf{$p$-value} & \textbf{Homogeneity} &  \\ \hline
\textbf{Speech Procduction} 
& \offtext & 0.961 & 0.005 & False & 1.340 & 0.261 & True & Kruskal-Wallis \\
 & \ontext & 0.961 & 0.005 & False & 1.340 & 0.261 & True & Kruskal-Wallis \\
  & \hopping & 0.966 & 0.012 & False & 1.323 & 0.267 & True & Kruskal-Wallis \\
 & \hopping{} \distal & 0.946 & $< 0.001$ & False & 0.419 & 0.794 & True & Kruskal-Wallis \\
 & \hopping{} \local & 0.945 & $< 0.001$ & False & 0.276 & 0.893 & True & Kruskal-Wallis \\
 & \sustained & 0.859 & $< 0.001$ & False & 1.539 & 0.197 & True & Kruskal-Wallis \\
 & \sustained{} \distal & 0.715 & $< 0.001$ & False & 1.203 & 0.315 & True & Kruskal-Wallis \\
 & \sustained{} \local & 0.791 & $< 0.001$ & False & 0.378 & 0.824 & True & Kruskal-Wallis \\
 \hline
\textbf{Reviewing} 
 & \offtext & 0.395 & $< 0.001$ & False & 1.025 & 0.412 & True & Kruskal-Wallis \\
 & \ontext & 0.395 & $< 0.001$ & False & 1.025 & 0.412 & True & Kruskal-Wallis \\
 & \hopping{} \ontext & 0.987 & 0.210 & True & 1.160 & 0.332 & True & \textbf{ANOVA} \\
 & \sustained{} \ontext & 0.876 & $< 0.001$ & False & 1.473 & 0.192 & True & Kruskal-Wallis \\ 
 & \regressionbt & 0.890 & $< 0.001$ & False & 0.524 & 0.789 & True & Kruskal-Wallis \\
 & \regressionin & 0.916 & $< 0.001$ & False & 2.486 & 0.026 & False & Kruskal-Wallis \\ 
 & \fc & 0.989 & 0.368 & True & 0.830 & 0.549 & True & \textbf{ANOVA} \\
 & \ffd & 0.990 & 0.412 & True & 0.662 & 0.680 & True & \textbf{ANOVA} \\
 & \gd & 0.990 & 0.412 & True & 0.662 & 0.680 & True & \textbf{ANOVA} \\
 & \ttt & 0.946 & $< 0.001$ & False & 0.658 & 0.683 & True & Kruskal-Wallis \\ \hline
\end{tabular}
}
\end{table*}
\begin{table*}[htbp]
  \centering
  \caption{Comparison between \emph{Speaking} and \emph{Reviewing} with Mann-Whitney U tests on hopping ontext ratio, sustained ontext ratio and off text ratio across five \interface{} conditions}
  \resizebox{\textwidth}{!}{
    \begin{tabular}{clrrrrrlrr}
    \toprule
    \multicolumn{1}{l}{Measure} & Condition & \multicolumn{1}{l}{Speaking Mean} & \multicolumn{1}{l}{Speaking SE} & \multicolumn{1}{l}{Reviewing Mean} & \multicolumn{1}{l}{Reviewing SE} & \multicolumn{1}{l}{U} & p     & \multicolumn{1}{l}{Z} & \multicolumn{1}{l}{Effect Size (r)} \\
    \midrule
    \multirow{5}[2]{*}{\hopping{}} & \plain{} & 0.302  & 0.040  & 0.341  & 0.026  & 180   & \multicolumn{1}{r}{0.598 } & -0.541  & -0.086  \\
          & \aoc{}   & 0.337  & 0.045  & 0.320  & 0.024  & 209   & \multicolumn{1}{r}{0.818 } & 0.243  & 0.038  \\
          & \rake{}  & 0.359  & 0.048  & 0.335  & 0.023  & 214   & \multicolumn{1}{r}{0.715 } & 0.379  & 0.060  \\
          & \gptsm{} & 0.361  & 0.053  & 0.340  & 0.022  & 209   & \multicolumn{1}{r}{0.818 } & 0.243  & 0.038  \\
          & \as{}    & 0.345  & 0.037  & 0.325  & 0.029  & 223   & \multicolumn{1}{r}{0.543 } & 0.622  & 0.098  \\
    \midrule
    \multirow{5}[2]{*}{\sustained{}} & \plain{} & 0.098  & 0.027  & 0.659  & 0.026  & 0     & <0.001 & -5.410  & -0.855  \\
          & \aoc{}   & 0.071  & 0.014  & 0.680  & 0.024  & 0     & <0.001 & -5.410  & -0.855  \\
          & \rake{}  & 0.073  & 0.012  & 0.665  & 0.023  & 0     & <0.001 & -5.410  & -0.855  \\
          & \gptsm{} & 0.108  & 0.023  & 0.660  & 0.022  & 0     & <0.001 & -5.410  & -0.855  \\
          & \as{}    & 0.096  & 0.023  & 0.675  & 0.029  & 0     & <0.001 & -5.410  & -0.855  \\
    \midrule
    \multirow{5}[2]{*}{\offtext{}} & \plain{} & 0.612  & 0.051  & 0.137  & 0.063  & 363   & <0.001 & 4.409  & 0.697  \\
          & \aoc{}   & 0.598  & 0.051  & 0.060  & 0.011  & 400   & <0.001 & 5.410  & 0.855  \\
          & \rake{}  & 0.578  & 0.050  & 0.064  & 0.016  & 396   & <0.001 & 5.302  & 0.838  \\
          & \gptsm{} & 0.537  & 0.064  & 0.047  & 0.008  & 400   & <0.001 & 5.410  & 0.855  \\
          & \as{}    & 0.564  & 0.039  & 0.114  & 0.048  & 380   & <0.001 & 4.869  & 0.770  \\
    \bottomrule
    \end{tabular}%
    }
  \label{tab:speak_vs_review}%
\end{table*}%

\begin{table*}[htbp]
  \centering
  \caption{Mean and Standard Error for Fixation metrics FC, FFD and TT for each \interface{}.} 
  \resizebox{\textwidth}{!}{%
    \begin{tabular}{clrrrrrrrrr}
    \toprule
          &       & \multicolumn{3}{c}{FC} & \multicolumn{3}{c}{FFD} & \multicolumn{3}{c}{TT} \\
\cmidrule{3-11}          &       & \multicolumn{1}{l}{Mean} & \multicolumn{1}{l}{Std Error} & \multicolumn{1}{l}{Sample size} & \multicolumn{1}{l}{Mean} & \multicolumn{1}{l}{Std Error} & \multicolumn{1}{l}{Sample size} & \multicolumn{1}{l}{Mean} & \multicolumn{1}{l}{Std Error} & \multicolumn{1}{l}{Sample size} \\
    \midrule
    \multirow{7}[2]{*} & \plain{} & 0.94  & 0.02  & 5839  & 134.20  & 1.96  & 5839  & 213.68  & 3.87  & 5839 \\
          & \aoc{}   & 0.88  & 0.01  & 6169  & 130.23  & 1.82  & 6169  & 194.70  & 3.31  & 6169 \\
          & \rake{}  & 0.92  & 0.01  & 5777  & 132.01  & 1.83  & 5777  & 202.09  & 3.72  & 5777 \\
          & \gptsm{} & 0.93  & 0.02  & 5392  & 130.21  & 1.85  & 5392  & 202.66  & 3.69  & 5392 \\
          & \as{}    & 1.05  & 0.02  & 3360  & 148.60  & 2.56  & 3360  & 236.88  & 4.97  & 3360 \\
          
          
    \bottomrule
    \end{tabular}%
    }
  \label{tab:token_level_metrics}%
\end{table*}%

\begin{table*}[htbp]
  \centering
  \caption{Reading metrics comparison between self-spoken text and external texts. }
  \resizebox{\textwidth}{!}{
    \begin{tabular}{clrcrrll}
    \toprule
    \textbf{Metric} & \multicolumn{1}{c}{\textbf{Test}} & \multicolumn{1}{c}{\textbf{H}} & \textbf{p} & \multicolumn{1}{c}{\textbf{$\eta^2$ or r}} & \multicolumn{1}{c}{\textbf{\combined{}}} & \multicolumn{1}{c}{\textbf{\easy{}}} & \multicolumn{1}{c}{\textbf{\hard{}}} \\
    \midrule
    \multirow{4}[2]{*}{\emph{FC}} & Kruskal-Wallis & 151.54  & <0.001 & 0.004  & \multicolumn{1}{l}{\meanse{0.935}{0.007}} & \meanse{0.952}{0.016} & \meanse{1.159}{0.019} \\
          & \combined{} vs \easy{} dunn's &       & 0.016 & -0.015  & \multicolumn{1}{l}{} &  &  \\
          & \combined{} vs \hard{} dunn's &       & <0.001 & -0.065  & \multicolumn{1}{l}{} &       &  \\
          & \easy{} vs \hard{} dunn's &       & <0.001 & -0.074  &       &  &  \\
    \midrule
    \multirow{4}[2]{*}{\emph{FFD}} & Kruskal-Wallis & 152.99  & <0.001 & 0.004  & \multicolumn{1}{l}{\meanse{133.812}{0.879}} & \meanse{145.738}{2.292} & \meanse{164.762}{2.536} \\
          & \combined{} vs \easy{} dunn's &       & <0.001 & -0.031  & \multicolumn{1}{l}{} &  &  \\
          & \combined{} vs \hard{} dunn's &       & <0.001 & -0.064  & \multicolumn{1}{l}{} &       &  \\
          & \easy{} vs \hard{} dunn's &       & <0.001 & -0.049  &       &  &  \\
    \midrule
    \multirow{4}[2]{*}{\emph{TT}} & Kruskal-Wallis & 224.02  & <0.001 & 0.006  & \multicolumn{1}{l}{\meanse{207.444}{1.714}} & \meanse{219.947}{4.168} & \meanse{285.602}{5.762} \\
          & \combined{} vs \easy{} dunn's &       & <0.001 & -0.025  & \multicolumn{1}{l}{} &  &  \\
          & \combined{} vs \hard{} dunn's &       & <0.001 & -0.081  & \multicolumn{1}{l}{} &       &  \\
          & \easy{} vs \hard{} dunn's &       & <0.001 & -0.083  &       &  &  \\
    \midrule
    \multirow{4}[2]{*}{\regressionin{}} & Kruskal-Wallis & 15.76  & 0.000  & 0.100  & \multicolumn{1}{l}{\meanse{13.730}{0.760}} & \meanse{7.900}{0.804} & \meanse{10.350}{1.334} \\
          & \combined{} vs \easy{} dunn's &       & 0.009  & 0.335  & \multicolumn{1}{l}{} &  &  \\
          & \combined{} vs \hard{} dunn's &       & 0.467  & 0.190  & \multicolumn{1}{l}{} &       &  \\
          & \easy{} vs \hard{} dunn's &       & 1.000  & -0.195  &       &  &  \\
    \bottomrule
    \end{tabular}%
    }
  \label{tab:interface_external}%
\end{table*}%

\clearpage
\section{System Prompt}

\subsection{Prompt for \gptsm{}}

\begin{quote}
\textbf{System: } 

Step 1: Please extract the core content of each sentence in the """content""" and bold it. You can bold the subject, verb or object phrases, and key modifiers/terms of the main sentence of the text below, and leave the rest of the text as it is without the key or original text. Be careful not to bold everything, """only bold {less than 60\%} of the text""" and output only the result, not the rest.  And use the following bold annotation method: <k>content</k>.
    
"""Please study and refer to the following example, and mark based on my output. All marked content should form complete sentences.""" For example: <k>The</k> recent <k>recognition of a link between</k> increasing rates of <k>deforestation and</k> increasing <k>global</k> climatic <k>warming has focused</k> new <k>attention on</k> the ecological role of <k>forests.</k>
Output the result wrapped in <new></new>.\\

Step 2: Refer to the past summary section. Output the newly added text with corresponding <k></k> tags, wrapped in <add></add>.\\

The purpose of Step 2 and Step 3 is to make sure you don't edit the <k></k> tags in past summary, and make sure the text stay true to the """content""" only add tags.\\

!!!Import note that the format of the output is :
"""<new>improve result</new>
<add>newly added content</add>
<summary>complete summary</summary> """ \\

\textbf{User: } "Past Summary: [last\_response] + Content: [\aoc{}\_result]"

\end{quote}

\subsection{Prompt for \as{}}

\begin{quote}
\textbf{System: } 

Please remove verbosity, repetition, intonation, and other irrelevant content from the following texts, optimise the presentation, fix speech recognition errors, fix grammatical errors, and then summarise and reconstruct the texts, trying to retain the meaning of the original presentation, as well as the consistency of the wording. Note that retain the same narrative perspective, and only the improved result is output, not other content.\\

Please directly output the complete summary without naming any sections. The result should not contain multiple parts but only the text of the summary. \\

\textbf{User: } "[\aoc{}\_result]"

\end{quote}
\end{document}